\documentclass[tighten, twocolumn, twocolappendix]{aastex631}
\usepackage{amsfonts} 
\usepackage{amsmath}
\usepackage{amssymb}
\usepackage{amsthm}
\usepackage{braket}
\usepackage{bm}
\usepackage{cancel}
\usepackage{color}

\usepackage{diagbox}
\usepackage{epsfig}
\usepackage{esint}
\usepackage{graphicx}
\usepackage{gensymb}
\usepackage{mathrsfs}
\usepackage{mathtools}
\usepackage{slashed}

\usepackage{xcolor}
\usepackage{tikz}
\usepackage{soul}

\usepackage{verbatim}

\tikzstyle{process} = [rectangle, minimum width=3cm, minimum height=1cm, text centered, draw=black, fill=gray!30, text width=3.0cm, rounded corners]
\tikzstyle{arrow} = [thick,->,>=stealth]

\begin{document}

\title{Probing Axions with Relativistic Jet Polarimetry}
\author[0009-0006-8832-2697]{Dashon Michel Jones}
\affiliation{%Rice University
Physics $\&$ Astronomy Department, Rice University, Houston,
Texas 77005-1827, USA}
\affiliation{Black Hole Initiative at Harvard University, 20 Garden Street, Cambridge MA 02138, USA}
\affiliation{Harvard-Smithsonian Center for Astrophysics, 60 Garden Street, Cambridge MA 02138, USA}

\author{Richard Anantua}
\affiliation{Physics and Astronomy Department, University of Texas at San Antonio, 1 UTSA Circle, San Antonio, TX 78249}
\affiliation{%Rice University
Physics $\& $Astronomy Department, Rice University, Houston,
Texas 77005-1827, USA}
\affiliation{Black Hole Initiative at Harvard University, 20 Garden Street, Cambridge MA 02138, USA}
\affiliation{Harvard-Smithsonian Center for Astrophysics, 60 Garden Street, Cambridge MA 02138, USA}

\author{Razieh Emami}
\affiliation{Harvard-Smithsonian Center for Astrophysics, 60 Garden Street, Cambridge MA 02138, USA}

\author{Nate Lujan}
\affiliation{Physics and Astronomy Department, University of Texas at San Antonio, 1 UTSA Circle, San Antonio, TX 78249}

\date{July 2025}

\begin{abstract}
The prospect of identifying axion signals due to axion-photon coupling induced changes to the polarization has now become a reality in view of near-horizon polarimetric observations by the Event  Horizon Telescope (EHT). Axion-like particles (ALPs), motivated as dark matter candidates by the strong CP problem, induce frequency-independent birefringence in linearly polarized radiation, producing observable rotations of the electric vector position angle.  
While previous studies have focused exclusively on axion signatures in near-horizon accretion disk emission, the relativistic jet of M87-- extending from 10 gravitational radii to kiloparsec scales-- remains unexplored as an axion probe despite offering extended path lengths through the putative dark matter distribution. In this study, we investigate the effects of an axion cloud around the jet in M87 on the Stokes maps of relativistic jets using a stationary, axisymmetric, self-similar model for the jet and a coherent, homogeneous soliton core in M87's galactic center for the axion background. At 230 GHz, for representative couplings in range $g_{a \gamma} \sim 5 \times 10^{-15} - 5 \times 10^{-14} GeV^{-1}$, we find that axion masses in the $10^{-21} eV $ range produce degree-level to multi-degree EVPA rotations, in some cases exceeding typical EHT measurement uncertainties, whereas masses in the $10^{-22} eV$ range yield predominantly sub-degree rotations. We identify a suite of morphological diagnostics that together constitute a framework for distinguishing axion-induced birefringence from plasma Faraday rotation in resolved jet polarimetry.
\end{abstract}

\section{\textbf{Introduction}}

The Event Horizon Telescope (EHT), which ushered in the age of direct imaging of supermassive black holes  with the sources M87* in 2019 and Sgr A* in 2022 \citep{2019ApJ...875L...1E,2022ApJ...930L..12E}, opens a new era for probing physics in the extreme environment of a supermassive black hole (SMBH). The unprecedented 20$\mu$as  resolution has opened many avenues for exploring fundamental physics with the EHT \citep{ayzenberg2023fundamentalphysicsopportunitiesnextgeneration}. In addition to facilitating tests of general relativity, accretion disk physics, and extreme electrodynamics in the strong gravity regime, EHT offers new avenues to study Beyond Standard Model (BSM) physics around compact objects. More recently, the EHT has been able to resolve the polarized images of both Sgr A* \citep{2024ApJ...964L..26E} and M87{\color{black}*} \citep{Akiyama_2021}, which has raised the interest of researchers looking to study BSM phenomenology around black holes. With the advantages of Very Long Baseline Interferometry (VLBI) and polarimetric measurements of the radiation, the EHT serves as a unique probe to constrain the existence of axion-like particles (ALPs) in \textcolor{black}{polarized} astrophysical environments.  Many of these searches typically attempt to search for axion effects on the electromagnetic sector \citep{Tobar_2019}. 
\par 
The axion is a hypothetical BSM particle motivated by being a solution to the strong CP problem \citep{PhysRevLett.40.223} in quantum chromodynamics (QCD). The QCD Lagrangian permits a CP-violating term, and yet the strong interaction appears to preserve CP symmetry Beyond QCD, ALPs generically appear in fundamental theories and string theories, serving as a viable dark matter candidate. There have been many proposed ways to search for axion signatures including nuclear magnetic resonance \citep{PhysRevLett.113.161801}, neutron stars \citep{noordhuis2024axioncloudsneutronstars}, and gravitational wave signatures \citep{Gorghetto_2021}.  Our methodology involves particularly distinctive, directly observable electromagnetic features. . 
\par
% Axions, and more generally, axion-like particles (ALPs), are prime dark matter candidates motivated by the Strong CP problem. 
Propagation of a linearly polarized photon in an ALP field induces periodic oscillations about the electric vector position angle (EVPA) \textcolor{black}{due to axion-photon coupling $g_{a\gamma\gamma}$. Ultralight axions may be treated as a classical wave due to their exceptionally high occupation number \citep{Hui_2021}. An analytic model of a DM halo formed in this regime can be described by a Schrodinger-Poisson system \citep{manita2024solitonselfgravitycorehalorelation} and a soliton core can form in the galactic center due to the balance between gravitational interactions and the quantum pressure \citep{Mocz_2023}. {\color{black} The soliton  (standing wave) core can also be used as a novel explanation for the presence of an observed rising proper motion for stars at the center of globular cluster 47-Tucanae \citep{2020PhRvD.101f3006E}. }
The spatial mass and density distribution of such halos is often described by a Navarro–Frenk–White (NFW) profile which is used in many cosmological simulations of cold dark matter (CDM) in galaxies, such as effects on structure formation \citep{Mina_2022}.  \color{black}{A dedicated recent comparison between different DM profiles, including CDM, Warm dark matter (WDM), and fuzzy-dark matter, showed that the morphologies of early galaxies are very sensitive to smoothness of their underlying filament network, providing a novel constraint on the nature of DM \citep{2025NatAs.tmp..246P}. }}
\par
When the Compton wavelength of the axion field is comparable to the size of the BH, the axion cloud can be excited due to rotational superradiance \citep{PhysRevLett.29.1114}. This is a generalized version of the Penrose process \citep{Lasota_2014} which typically focuses on a classical particle \citep{Penrose1971ExtractionOR}. This configuration is also often referred to as the gravitational atom, since the energy eigenstates are similar to those of the hydrogen atom \citep{Baumann_2020}. Historically, searches for BH superradiance signatures have focused on gravitational wave signals from bosenova \citep{Arvanitaki_2011}. More recently, it has been proposed that the high spatial resolution and polarimetric measurements of the EHT can be used to probe axion cloud around an SMBH \citep{PhysRevLett.124.061102}. This is because a photon passing through an axion cloud can have axion-induced time oscillations of the Electric Vector Position Angle (EVPA), similar to the classical Faraday rotation effect. 
\par
Known since Heber Curtis observed a ``curious" straight ray in the giant elliptical galaxy M87 in 1918, relativistic jets are powerful particle and radiation outflows emanating from a wide range of black holes throughout the Universe. They are the most energetic discrete astrophysical sources- with the blazar jet $\rm S0528+134$ achieving a remarkable bolometric luminosity of $10^{49}$ erg/s \citep{Volvach2024}. Jets co-develop with their host galaxies, as they are fed by black hole accretion of magnetized plasma and are responsible for important galactic feedback processes \citep{Cielo2018}.  The leading mechanisms for powering jets include the extraction of rotational energy from the ergoregion of rotating black holes by vertical magnetic field lines \citep{Blandford1977} and the production of hydromagnetic winds from accretion disks \citep{Blandford1982}. The former mechanism is supported by Event Horizon Telescope polarized M87{\color{black}*} and Sgr A* observations of spiral EVPA patterns on the emitting ring around supermassive black holes. Thus, in what follows, we model our relativistic jet testbed for axion signatures using the Blandford-Znajek mechanism. 
\par
The axion-EHT literature has concentrated almost entirely on accretion disk emission and photon ring polarization patterns. The work of \cite{PhysRevLett.124.061102} demonstrated that the polarimetric measurements of M87* could constrain ultralight axion-like particles in the mass range of $10^{-21}$ to $10^{-20}$ eV\textcolor{black}{.}  
M87's relativistic jet-extending from $\rm \sim 10\,r_g$ to
\textcolor{black}{tens of thousands of parsecs} remains unexplored as an axion probe despite several potential advantages. Jets offer extended path lengths through the putative axion field, sampling the dark matter distribution across galactic scales rather than only the horizon region. 
\par
In this paper, we propose searches for axions around M87{\color{black}*} using polarimetric measurements of relativistic jets. We posit that relativistic jets can be used as another laboratory for studying ALPs and dark matter in conjunction with horizon scale accretion flow. The analysis focuses on how ALPs could theoretically affect the polarization maps of jets. 
\par
This paper is structured as follows. In Section 2, we present the configuration for an ambient DM cloud that forms a soliton core in the galactic center of M87. This is well motivated and is the simplest to work with. Section 3 reviews the jet model found in \cite{Anantua_2020} followed by Section 4 which discusses the interaction between axions and the Standard Model, particularly electromagnetism. In Section 5, we present methods used to study the polarization maps with and the axion-induced maps generated for the soliton core model. We summarize in Section 6 and present further analyses to be undertaken in the near future. This paper develops a morphological framework for characterizing axion-induced signatures, establishing simple physical intuition for constructing future templates. As such, the paper does not aim to place parameter-space constraints which is deferred for future work. 

\section{\textbf{Coherent ALP Field}}

The dark matter model used here is very similar to that seen in \cite{Yuan._2021}. A linearly polarized photon traveling through an ALP field can have periodic rotation of the EVPA. The relevant Lagrangian for an ALP field coupled to electromagnetism is the following:
\begin{equation}
    \mathcal{L} = -\frac{1}{4}F_{\mu\nu}F^{\mu\nu} + \frac{1}{2}\partial_\mu a\,\partial^\mu a - \frac{1}{2}m_a^2\,a^2 - \frac{g_{a\gamma}}{4}\,a\,F_{\mu\nu}\tilde{F}^{\mu\nu}
\end{equation}
where $F_{\mu \nu} = \partial_\mu A_{\nu} - \partial_\nu A_\mu$ and $\tilde{F}^{\mu\nu}$ are the dual. The equation of motion for the ALP field arising from this is,
\begin{equation}
    \square a + m_{a}^2 = g_{a \gamma} \boldsymbol{E \cdot} \boldsymbol{B}
\end{equation}
By assuming that the source term is negligible compared to $m_a^2 a$, we can ignore the back-reaction term. The solution then becomes a coherently oscillating ALP field,
\begin{equation}
    a(t,\vec{x}) \approx a_{0} \sin[m_at - \delta(\vec{x})], \quad
    a_0 = \sqrt{\frac{\rho_{DM}}{m_a^2}}
\end{equation}
where $\delta(\vec{x})$ is a position-dependent phase. For axion experiments on Earth, the value of this tends to be in the vicinity of $\rho_{DM} \approx 0.4 GeV / cm^{-3}$. This value tends to be much larger in many astrophysical systems, including near a Kerr black hole. It is valid to describe such a field as a classical field since at masses $<30\rm eV$, the de Broglie wavelength is much greater than the galaxy interparticle separation and the average number of particles in the de Broglie volume is large enough such that the axion cloud behaves like a Bose-Einstein Condensate \citep{Hui_2021}. For ultralight axion DM, the balance between quantum pressure and galactic interactions can give rise to a soliton core near the galactic center. The radius of the core is directly related to the de Broglie wavelength of the ALP field,
\begin{equation}
    r_c = \frac{2 \pi}{m_a v} 
\end{equation}
where $v$ is the average velocity of the dark matter within the specified galaxy. Within the solitonic core, the dark matter density is assumed to be coherent and homogeneous whereas outside the soliton core, the DM density follows a NFW profile. Simulations from \cite{Murphy_2011} suggest the mean velocity of dark matter in M87 to be around $800\ \rm {km}/{s}$ corresponding to $v \approx 2.66 \times 10^{-3} \rm c$. The relationship between mass and core radius can be found in Fig. 1.

\begin{figure}[!h] 
        \centering \includegraphics[width=0.95\columnwidth]{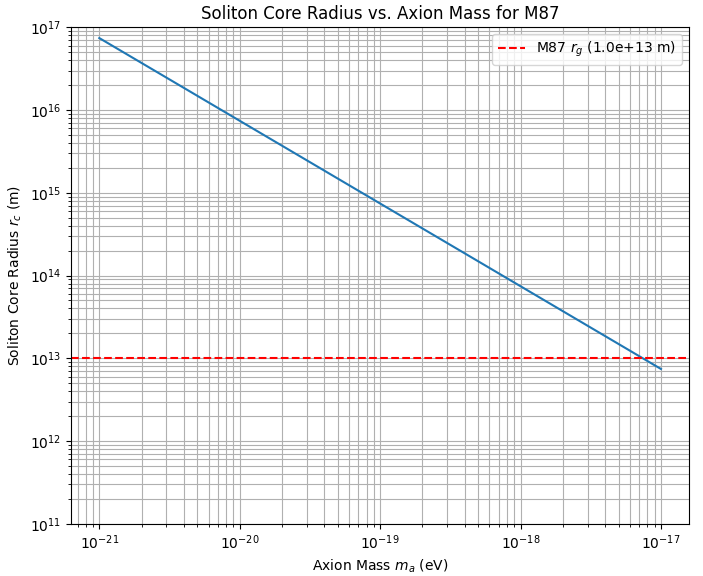}
        \caption{ Relationship between axion mass and soliton core radius.}
        \label{fig:SolitonRadius}
\end{figure}

Simulations from \cite{Schive_2014} show that the density profile can be parametrized as follows, 

\begin{equation}
\rho_{\mathrm{DM}}= \begin{cases}0.019 \times\left(\frac{m_{a}}{10^{-22} \mathrm{eV}}\right)^{-2}\left(\frac{r_{c}}{1 \mathrm{pc}}\right)^{-4} \left(\frac{M_{\odot}} {\mathrm{pc}^{3} }\right), & \text { for } r<r_{c}  \\ \frac{\rho_{0}}{r / R_{g}\left(1+r / R_{g}\right)^{2}}, & \text { for } r>r_{c}\end{cases}
\end{equation}

\begin{figure}[!h] 
        \centering \includegraphics[width=0.90\columnwidth]{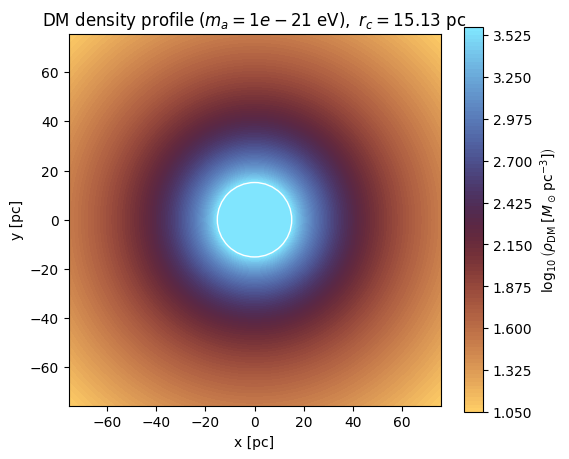}
        \caption{ Density profile for dark matter with axion mass of $10^{-21} \rm eV.$ White circle highlights the radius of the soliton core of which, density falls off as NFW profile outside this. The black hole would lie in the center of the soliton with the soliton being much larger than the black hole and extension of the jet.}
        \label{fig:Density}
\end{figure}

Additionally, from the Lagrangian, we get the equation of motion for the electromagnetic field,

\begin{equation}
\ddot{\boldsymbol{A}}-\nabla^{2} \boldsymbol{A}=g_{a \gamma}(\dot{a} \nabla \times \boldsymbol{A}+\dot{\boldsymbol{A}} \times \nabla a) .
\end{equation}

The equations of motion result in two different dispersion relations for each of the two different circularly polarized modes. 

\begin{equation}
w_{ \pm}^{2}-k^{2} \mp g_{a \gamma}(\dot{a}+\hat{\boldsymbol{k}} \cdot \nabla a)|k|=0 
\end{equation}

The misalignment of the ALP field causes spontaneous breaking of the parity symmetry. Such an effect causes birefringence, and for a linearly polarized photon, we will see a change in the position angle in the polarization plane. 

\par
One caveat on this dark matter model is that the presence of a large of enough black hole in the galactic center should cause the soliton core to squeeze the soliton core, reducing its radius \citep{Mocz_2023}. For M87, a cosmological core–halo FDM soliton is not expected to persist for $m_a > 10^{-22} eV$. Accordingly, throughout this work we model the inner dark matter enhancement as a phenomenological axion or ALP overdensity used to quantify birefringent propagation signatures. The adopted soliton function should be interpreted as a convenient proxy for a compact axion density profile rather than a guaranteed cosmological FDM soliton in M87. A dedicated treatment of formation and survival in the SMBH environment, as well as effects on EVPA, is beyond the scope of this work. Instead we focus on the observable imprints on the polarization maps on the jets, conditional on the presence of a coherent ALP field.

\section{\textbf{ANALYTIC RELATIVISTIC JET MODEL}}
To investigate axion signatures in the polarized emission from M87's relativistic jet, we employ the semi-analytic jet model developed by \cite{Anantua_2020}. This model provides a computationally efficient framework for generating synthetic polarimetric observations while capturing the essential physics of magnetically dominated jet flows. The advantage of this approach is twofold: it enables us to explore parameter space systematically without the computational expense of a full general relativistic magnetohydrodynamic (GRMHD) simulation, and its self-similar structure allows us to model jet physics across a wide range of spatial scales-from the black hole horizon ($\sim10M$) to large-scale jet structures ($\sim10^5M$), where $M \equiv GM_{\rm BH}/c^2$ is the gravitational radius.

\subsection{Self-similar Stationary Semi-analytic Model}

The jet model is based on force-free regions of relativistic GRMHD simulations and assumes the plasma is stationary ($\partial/\partial t = 0$) and axisymmetric ($\partial/\partial \phi = 0$). These assumptions are well- justified for the time-averaged structure of AGN jets and significantly simplify the analysis while preserving the key electromagnetic properties relevant for polarization studies. The model is based on  a MAD $\rm a/M=0.9$ GRMHD simulation jet region in which the force-free  jet boundary is clearly delineated from the turbulent plasma in the MHD limit outside a parabolic interface \citep{Anantua2016PhD}. This prototypical Blandford-Znajek jet has a ``spine" of relativistic plasma, where $v_z$ accelerates along the axis surrounded by a non-relativistic sheath. 

The model employs cylindrical coordinates $(s, \phi, z)$, where $s$ is the cylindrical radius and $z$ is the height along the jet axis. The self-similarity of the flow is characterized by the dimensionless variable:
\begin{equation}
\xi = \frac{s^2}{z}
\end{equation}

This self-similarity allows the magnetic flux $\Phi$, current $I$, and field-line angular velocity $\Omega_B$ to be expressed as functions of $\xi$ alone, rather than $s$ and $z$ independently. The magnetic field components are then derived from the flux function through:
\begin{equation}
B_s = -\frac{1}{2\pi s}\frac{\partial \Phi}{\partial z}, \quad B_z = \frac{1}{2\pi s}\frac{\partial \Phi}{\partial s}
\end{equation}
with the toroidal component given by:
\begin{equation}
B_\phi = \frac{I(\xi)}{2\pi s}
\end{equation}

For a force-free plasma, where the Lorentz force vanishes ($\rho \mathbf{E} + \mathbf{j} \times \mathbf{B} = 0$), the current satisfies:
\begin{equation}
I(\xi) = -\frac{2\Omega_B}{\Phi'}
\end{equation}
where primes denote derivatives with respect to $\xi$.

The specific functional forms for $\Phi(\xi)$ and $\Omega_B(\xi)$ are obtained by fitting to azimuthally averaged quantities from GRMHD simulations of a magnetically arrested disk around a spinning black hole \citep{McKinney_2012}. \cite{Anantua_2020} adopts the fitting functions:
\begin{equation}
\Phi(\xi) = \tanh(0.3\xi)
\end{equation}
\begin{equation}
\Omega_B(\xi) = 0.15\exp(-0.3\xi^2)
\end{equation}
which capture the essential structure of the jet: a highly magnetized spine near the axis (small $\xi$) surrounded by return currents at larger radii (larger $\xi$). The velocity field includes both poloidal components derived from the electromagnetic structure and a toroidal component from field-line rotation. This novel formalism,  has been used to study the horizon scale polarimetric signatures of positrons in M87* \citep{2020ApJ...896...30A,2021ApJ...923..272E}.
\par
For application to M87, the model is scaled using a black hole mass of $M_{\rm BH,M87} = 6.6 \times 10^9 \, M_\odot$, corresponding to a gravitational radius $r_g \approx 10^{13}$ m and light-crossing time $\sim9$ hours. The magnetic flux is normalized to $\Phi_{z,\rm M87} \approx 10^{26}$ Wb, consistent with estimates for M87's jet. The viewing angle is set to $\theta = 20^\circ$, matching observational constraints for M87.

\subsection{Constant $\beta_e$ Model for Plasma Emission}

To convert the electromagnetic structure into observable synchrotron emission, we require a prescription for the particle energy distribution. Various different emission models have been proposed in the literature \citep{2024MNRAS.528..735A}. Here we employ a ``constant $\beta_e$" prescription, where the partial pressure $\tilde{P}_e$ of synchrotron-emitting relativistic electrons and positrons is related to the local magnetic pressure:
\begin{equation}
\tilde{P}_e = \beta_{e0} P_B = \beta_{e0} \frac{B^2}{2}
\end{equation}

Here $\beta_{e0}$ is a dimensionless parameter representing the ratio of particle pressure to magnetic pressure. This formulation is motivated by the expectation that a fraction of the electromagnetic energy density is converted to particle energy through processes such as magnetic reconnection, shocks, or turbulent acceleration.

The sub-equipartition regime ($\beta_{e0} \ll 1$) is expected for the highly magnetized inner jet regions that dominate millimeter-wavelength emission. Based on observational constraints from M87's spectrum and jet geometry, characteristic values are $\beta_{e0} \sim 10^{-10}$ to $10^{-6}$, with lower values corresponding to more magnetized plasmas closer to the black hole.

The emitting particles are assumed to follow a power-law energy distribution:
\begin{equation}
N_e(\gamma) = K_e \gamma^{-p} \quad \text{for } \gamma_{\min} < \gamma < \gamma_{\max}
\end{equation}
where $\gamma$ is the particle Lorentz factor, $p$ is the spectral index (typically $p \approx 2.5-3$ for AGN jets), and $\gamma_{\min} = 10$ represents the minimum energy of relativistic particles contributing to synchrotron emission. This distribution captures the non-thermal particle acceleration processes operating in relativistic jets.

For plasmas containing a mixture of electrons, positrons, and protons/ions, the model can be generalized to account for composition. If we denote the ion number density as $n_i$ and the total particle density as $n$, the effective emitting particle pressure becomes:
\begin{equation}
\tilde{P}_e = \frac{n - n_i}{n} \beta_{e0} P_B
\end{equation}

This modification reflects the fact that only electrons and positrons contribute significantly to synchrotron emission at observed frequencies-protons are too massive (by a factor of 1836) to emit efficiently at the same energies. In this work focused on the direct effect of axions on emission, thus, we set the positron fraction to 0.

\subsection{Polarized Radiative Transfer}

The polarized synchrotron emission is described using the Stokes parameters $(I, Q, U, V)$, which fully characterize the intensity $I$ and polarization state of the radiation. The evolution of these parameters along a photon trajectory is governed by the polarized radiative transfer equation:
\begin{equation}
\frac{d}{ds}\begin{pmatrix} I \\ Q \\ U \\ V \end{pmatrix} = \begin{pmatrix} j_I \\ j_Q \\ j_U \\ j_V \end{pmatrix} - \begin{pmatrix} \alpha_I & \alpha_Q & \alpha_U & \alpha_V \\ \alpha_Q & \alpha_I & \rho_V & -\rho_U \\ \alpha_U & -\rho_V & \alpha_I & \rho_Q \\ \alpha_V & \rho_U & -\rho_Q & \alpha_I \end{pmatrix} \begin{pmatrix} I \\ Q \\ U \\ V \end{pmatrix}
\end{equation}
where $s$ is the affine parameter along the ray path. The source vector components $j_I, j_Q, j_U, j_V$ represent polarized emission; $\alpha_I, \alpha_Q, \alpha_U, \alpha_V$ are absorption coefficients; and $\rho_V, \rho_Q, \rho_U$ are Faraday rotation and conversion coefficients.

The emission and absorption coefficients are computed from the synchrotron formalism for a power-law particle distribution. These coefficients depend on the effective magnetic field $B_e = |\mathbf{B} \times \hat{n}|$, where $\hat{n}$ is the line-of-sight direction, and on the partial pressure $\tilde{P}_e$ through the $\beta_{e0}$ parameter. The key dependencies are:
\begin{equation}
j_I \propto \beta_{e0} B_e^{(p+1)/2} \nu^{-(p-1)/2}
\end{equation}
\begin{equation}
\alpha_I \propto \beta_{e0} B_e^{(p+2)/2} \nu^{-(p+4)/2}
\end{equation}
with similar expressions for other Stokes components. The linear polarization components $Q$ and $U$ reflect the projected magnetic field orientation, while circular polarization $V$ arises from higher-order effects and asymmetry in the particle distribution.

As polarized light propagates through magnetized plasma, Faraday rotation causes the plane of linear polarization to rotate:
\begin{equation}
\frac{d\chi}{ds} = \rho_V
\end{equation}
where $\chi$ is the electric vector position angle (EVPA). The Faraday rotation coefficient is given by:
\begin{equation}
\rho_V \propto \frac{n_e B_\parallel}{\nu^2}
\end{equation}
where $B_\parallel$ is the magnetic field component parallel to the line of sight and $n_e$ is the electron density. Faraday conversion (characterized by $\rho_Q$ and $\rho_U$) couples circular and linear polarization modes.

For the plasma compositions considered, both electrons and positrons contribute to Faraday rotation, but with opposite signs due to their opposite charges. In a pure electron-positron plasma, these contributions partially cancel, while in electron-proton plasmas, only electrons contribute significantly.

The radiative transfer equation is solved by integrating along rays from the observer plane $(X, Y)$ backward through the jet to the far side. Special relativistic effects are incorporated through Doppler boosting factors that depend on the bulk fluid velocity and viewing angle. The observed frequency $\nu$ relates to the comoving frequency $\nu'$ through:
\begin{equation}
\nu = \mathcal{D} \nu', \quad \mathcal{D} = \frac{1}{\Gamma(1 - \vec{\beta} \cdot \hat{n})}
\end{equation}
where $\Gamma$ is the bulk Lorentz factor and $\vec{\beta}$ is the bulk velocity in units of $c$.

The investigations in \cite{Anantua_2020} produced synthetic Stokes maps at observing frequencies of 86 GHz and 230 GHz-- the two primary frequencies of Event Horizon Telescope observations. These maps reveal characteristic patterns: bilateral asymmetry in intensity $I$ due to Doppler beaming, symmetric/antisymmetric structures in $Q$ and $U$ reflecting the helical magnetic field geometry, and varying degrees of circular polarization $V$ depending on plasma composition and magnetization.

This framework provides the baseline polarized emission against which axion-induced modifications will be compared. As will be described in Section 4, the presence of an axion field introduces an additional contribution to the Faraday rotation coefficient, which alters the evolution of the Stokes parameters along photon trajectories and produces observable signatures in the polarization maps.

\section{\textbf{Axion Electrodynamics and Radiative Transfer}}
Since the EHT can resolve inflows and outflows around supermassive black holes, %electromagnetic signatures,
including the polarized images of M87, we consider how these signatures would change in the presence of an axion field. While modified Maxwell equations help integrate the dynamics of axion-electromagnetic interactions, for generating ray traced images, it is more important to couple the the field to the radiative transfer equations to see how the photon polarization evolves as it propagates through an ALP field. Here, we use the geometric optics approximation of photons propagating through an ALP field in curved spacetime \citep{Schwarz_2021}. The radiative transfer model is identical to \cite{Chen_2022} with main equations being pulled from it. 

\subsection{Modified Maxwell Equations}
When axions are coupled to electromagnetism, Maxwell's equations are modified in the presence of the field \citep{PhysRevLett.51.1415}. The modifications come from the interaction term in the Lagrangian.

\begin{equation}
    \mathcal{L} _{a \gamma} = \frac{g_{a \gamma}}{4}a F_{\mu \nu} \tilde{F}^{\mu\nu}
\end{equation}

The equations of motion for the axion and gauge fields can be written as,

\begin{equation}
    \begin{split}
    &\partial_t^2 a - \nabla^2 a + m_a^2\,a = -\frac{g_{a\gamma}}{4}\,F_{\mu\nu}\tilde{F}^{\mu\nu} \\
    &\partial_{\mu} F^{\mu \nu} = - j^{\nu}, \quad \partial_{\mu} \tilde{F}^{\mu\nu} = 0
    \end{split}
\end{equation}
where $j^{\nu} = g_{a \gamma} \partial_{\mu} \tilde{F}^{\mu\nu}$. From this, it is straightforward to derive the axion-modified Maxwell equations,

\begin{gather}
    \nabla \cdot \mathbf{E} = \rho\ -\ g_{a\gamma} \mathbf{B} \cdot \nabla a \\
    \nabla \times \mathbf{B} = \dot{\mathbf{E}}\ +\ \mathbf{J}\ +\ g_{a\gamma}\ \left(\ \dot{a}\ \mathbf{B} - \mathbf{E} \times \nabla a\ \right) \\
    \nabla \times \mathbf{E} = - \dot{ \mathbf{B}} \\
    \nabla \cdot \mathbf{B}  = 0
\end{gather}

Note, the axion-electromagnetic charge and current are defined as follows.

\begin{equation}
    \rho_a = - \ g_{a\gamma} \mathbf{B} \cdot \nabla a, \quad \mathbf{J}_a = g_{a\gamma}\ \left(\ \dot{a}\ \mathbf{B} - \mathbf{E} \times \nabla a\ \right)
\end{equation}

\subsection{Radiative Transfer with Axions}
For our radiative transfer analysis, we apply the geometric optics approximation. In the Lorenz gauge, the equation of motion for the electromagnetic field can be written,

\begin{equation}
\nabla_{\mu} \nabla^{\mu} A^{\nu}-R_{\nu}^{\mu} A_{\mu}=-g_{a  \gamma}\left(\nabla_{\mu} a\right) \tilde{F}^{\mu \nu} 
\end{equation}
which under the geometric optics approximation has the following ansatz,

\begin{equation}
    A_\mu(x)\sim \Re\!\bigl\{\,\bar A_\mu(x)\,e^{\,iS(x)/\varepsilon}\bigr\}.
\end{equation}
Here, we define the four-dimensional wave vector as $k_{\mu} \equiv {1}/({\epsilon}) \partial_{\mu} S(x)$. At leading order in $\varepsilon^{-1}$, photons follow null geodesics ($k^\mu k_\mu=0$), and we require this condition to be true along the path of the photons. At next order, the vector potential expansion yields, 

\begin{equation}
k^{\mu} \nabla_{\mu} \bar{A}^{\nu}+\frac{1}{2} \bar{A}^{\nu} \nabla_{\mu} k^{\mu}+g_{a \gamma} \epsilon^{\mu \nu \rho \sigma} \bar{A}_{\sigma} k_{\rho} \nabla_{\mu} a=0 . 
\end{equation}
This can be further simplified if we introduce a space-like polarization vector $\xi^\mu$ that satisfies $\xi_\mu k^\mu =0$. This allows us to decompose the vector potential equation into two equations, one for amplitude $\bar{A}$ and one for polarization vector.

\begin{gather}
k^{\mu} \nabla_{\mu} \bar{A}+\frac{1}{2} \bar{A} \nabla_{\mu} k^{\mu}  =0  \\
k^{\mu} \nabla_{\mu} \xi^{\sigma}+g_{a  \gamma} \epsilon^{\mu \nu \rho \sigma} k_{\mu} \xi_{\nu} \nabla_{\rho} a  =0
\end{gather}

Note that the equation of motion for $\bar{A}$ does not contain the axion field. This means that the intensity of the light is not affected by the presence of the field. Projecting onto circular polarization states and the reference frame of the observer, one can show that the left- and right-handed modes acquire opposite phase shifts proportional to the change in the axion field along the ray.  For a linearly polarized photon, this corresponds to a net rotation of the electric-vector position angle (EVPA),

\begin{equation}
    \Delta\chi \;\approx\; g_{a\gamma}\,\bigl[a(\text{observer}) - a(\text{emission})\bigr],
    \label{endpointBirifringence}
\end{equation}

\par
It is usually best, and most convenient, to describe photon propagation near the BH with the covariant, polarized, radiative transfer equations, solving for the Stokes vector $I = (I, Q, U, V )$. These include $j_S$, absorption $\alpha_S$, as well as Faraday rotation/conversion coefficients $\rho_V,\rho_Q,\rho_U$. As stated before, the intensity of the light is unaffected. The axion field can alter the Faraday conversion coefficient term. Splitting contributions from the plasma and the axions we can define the projection matrix,

\begin{equation}
M=M_{\mathrm{plasma}}+M_{\mathrm{axion}}
\end{equation}

where the first term is exactly the Muller Matrix in the ordinary radiative transfer equations,

\begin{equation}
M_{\text {plasma }} \equiv\left(\begin{array}{cccc}
\alpha_{I} & \alpha_{Q} & \alpha_{U} & \alpha_{V}  \\
\alpha_{Q} & \alpha_{I} & \rho_{V} & -\rho_{U} \\
\alpha_{U} & -\rho_{V} & \alpha_{I} & \rho_{Q} \\
\alpha_{V} & \rho_{U} & -\rho_{Q} & \alpha_{I}
\end{array}\right)
\end{equation}

The axion matrix is simply characterized as,

\begin{equation}
    M_{\mathrm{axion}}=\left(\begin{array}{cccc}
0 & 0 & 0 & 0  \\
0 & 0 & -2 g_{a \gamma \gamma} \frac{d a}{d s} & 0 \\
0 & 2 g_{a \gamma \gamma} \frac{d a}{d s} & 0 & 0 \\
0 & 0 & 0 & 0
\end{array}\right)
\end{equation}

Hence we get the following correction to the Faraday rotation,

\begin{equation}
    \rho_{V}^{\prime}=\rho_{plasma}-2 g_{a \gamma} \frac{d a}{d s} 
    \label{Differential EVPA}
\end{equation}

where s is the affine parameter of the photon trajectory. The derivative $\frac{da}{ds}$ can be interpreted as the change in the axion field as "seen" by the photon along its trajectory. This arises from the parallel transport of the polarization vector in the presence of the axion gradient(i.e. change in axion field along the photon's worldline). In other words, for ray-traced images, the axion effect can be included by a change to the Faraday rotation coefficient. This directly effects the EVPA since,

\begin{equation}
\chi \equiv \frac{1}{2} \arg (Q+i U) .
\end{equation}

\begin{figure*}[t]
    \centering
    \includegraphics[width=0.98\textwidth]{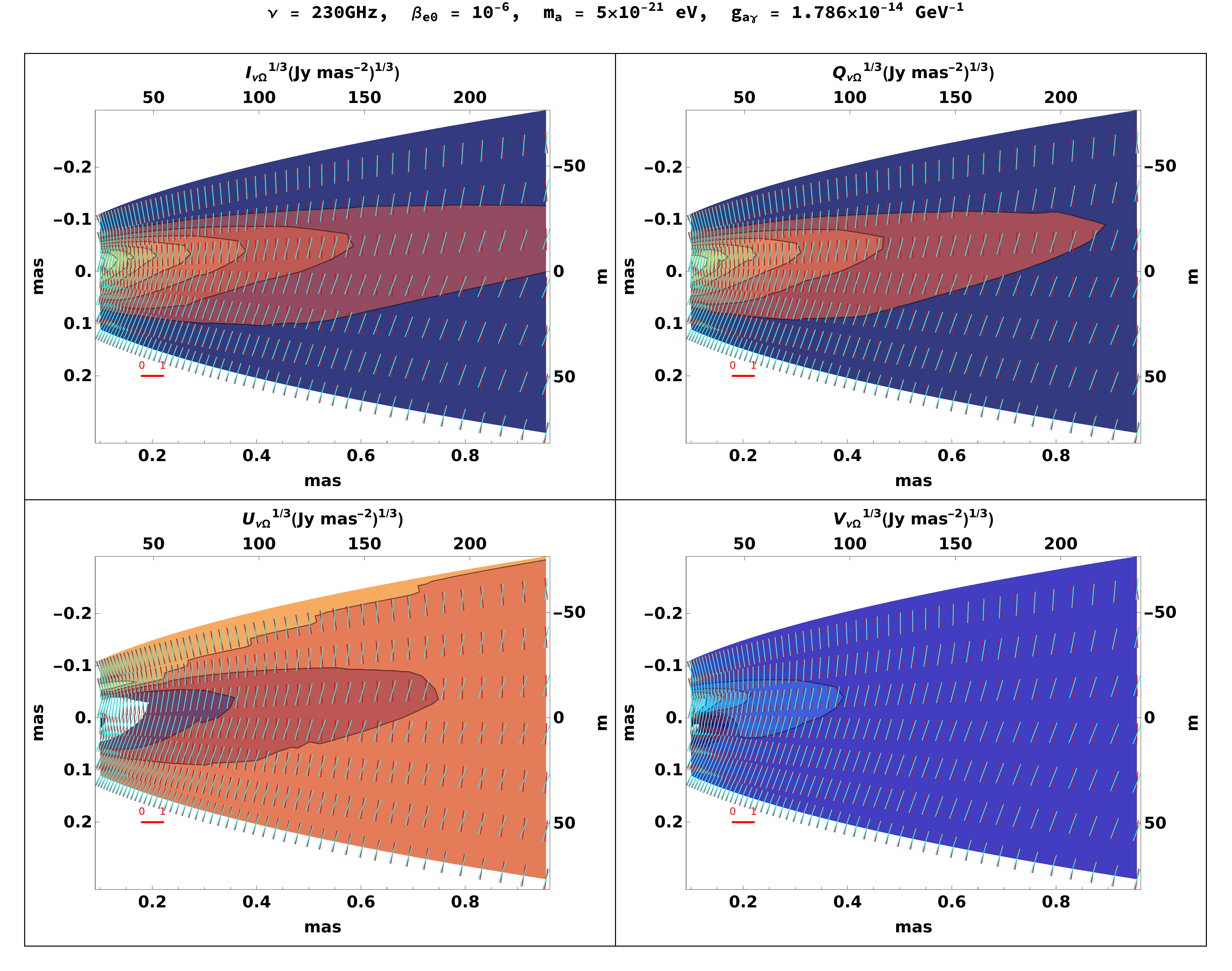}
    \caption{Synthetic Stokes maps for $m_a = 5 \times 10^{-21} eV$ and $g_{a \gamma} = 1.786 \times 10^{-14} GeV^{-1}$. The red EVPA ticks correspond to the EVPA angle without axions while the cyan-colored ticks correspond to the EVPA angle with axions. This example shows clearly the difference in EVPA angle when the jet is surrounded by the axion cloud vs. when it is not. }
    \label{fig:Stokes1}
\end{figure*}

Besides axion-photon interactions, the polarization state is affected by the medium effects, such as by the Faraday conversion and rotation, optical thickness etc. Hence, it's typically best to use the differential radiative transfer to properly describe birefringence effects that would be seen on the observer plane. It should also be noted that Eq. \eqref{endpointBirifringence} is an approximation for a single, linearly polarized photon in a vacuum. One should solve \eqref{Differential EVPA} for more realistic cases. Henceforth, we solve this equation to include medium effects in our radiative transfer pipeline.

\section{\textbf{Synthetic Stokes Maps}}

We now apply the radiative transfer framework of Sections 3-4 to generate synthetic polarimetric observations of M87's jet in the presence of an axion field. We compute Stokes maps at observing frequencies of 86 GHz and 230 GHz, matching EHT observational campaigns. Since our aim is general tomographic structure as opposed to parameter space constraints, the four mass values we analyze are $1 \times 10^{-22}\rm eV$, $5 \times 10^{-22}\rm eV$, $1 \times 10^{-21}\rm eV$ , and $5 \times 10^{-21}\rm eV$. Accordingly, we restrict ourselves to the couplings in the window of $5 \times 10^{-15} - 5 \times 10^{-14} \rm GeV^{-1}$. Since the focus is on axion effects, we restrict the plasma composition of the jet to be a purely symmetric ionic electron plasma  with a fixed $\beta_{e0} = 10^{-6}$. Further analysis for effects of plasma composition on Stokes maps can be found in \cite{Anantua_2020}. The jet region analyzed here is much smaller than the soliton core for each mass. We explicitly focus on the case t=0 and essentially show ``snapshots" of the EVPA angle change in the presence of an ALP field. While the ALP field induces time evolving EVPA angle changes, our focus is on the spatial morphology properties induced from the jet for further proxy. A full time dependence analysis is saved for the future. As such, this should be interpreted as the maximal change we expect to see in the EVPA, due to the axion field strength reaching its peak amplitude.

\subsection{Stokes Maps Properties}

\begin{figure*}[t]
    \centering
    \includegraphics[width=0.98\linewidth]{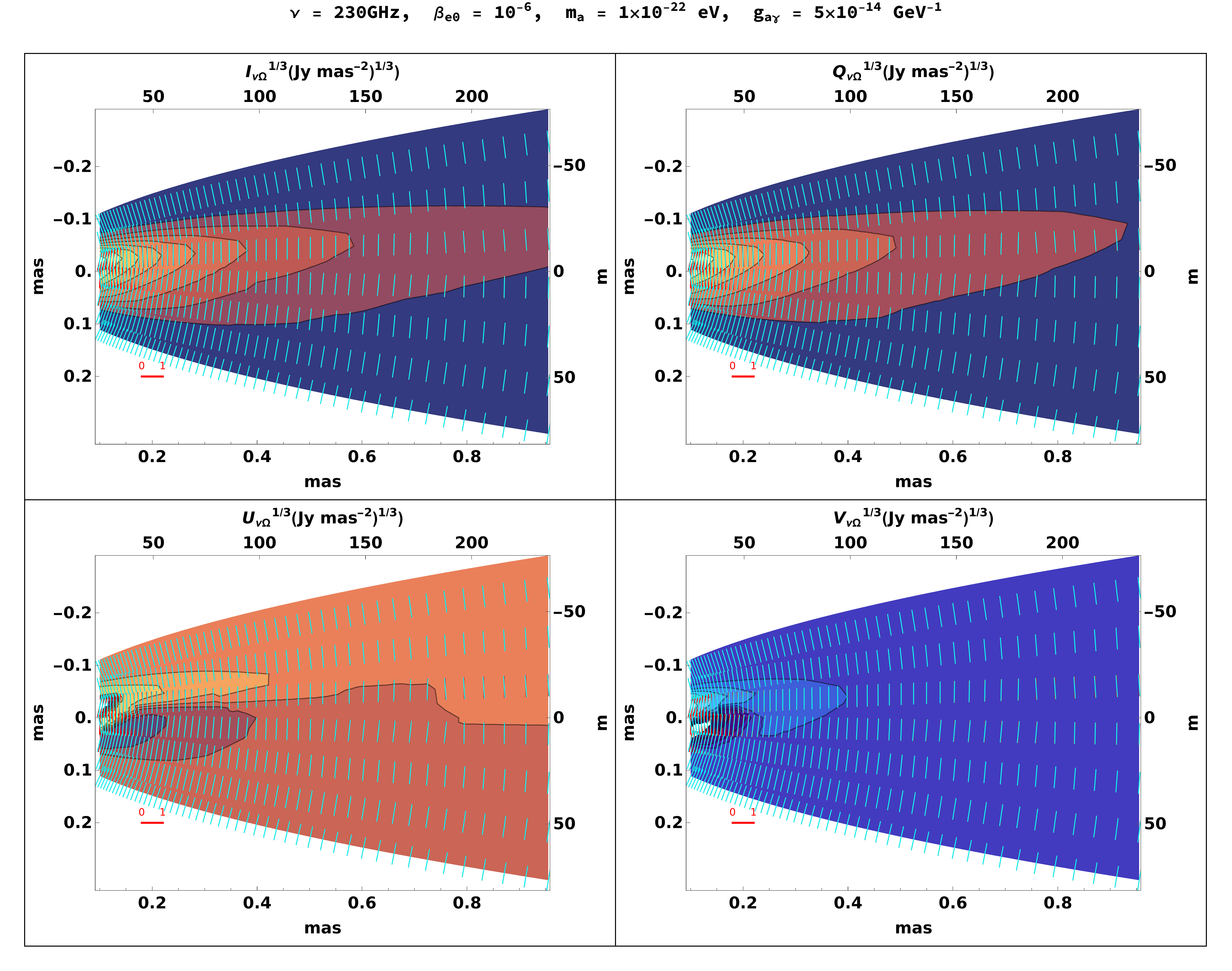}
    \caption{Example stokes map for $m_a = 10^{-22} eV$ and $g_{a \gamma} = 5 \times 10^{-14} GeV^{-1}$. The example is meant to highlight the weakness of the signal in comparison for this mass despite the strongest coupling in our dataset being used.  }
    \label{Stokes2}
\end{figure*}

 We note that given the timestamp of which these plots were taken, these represent the maximal EVPA change we expect to see.  While the Stokes parameter $Q$ and $U$ encode the changes in the EVPA angle, the EVPA vectors are shown on each map for convenience. Figure \ref{fig:Stokes1} presents representative Stokes maps for an axion mass of $m_a = 5 \times 10^{-21}$ eV and coupling constant $g_{a\gamma} = 1.786 \times 10^{-14}\rm GeV^{-1}$, observed at 230 GHz.   
The four panels show the standard Stokes parameters $(I, Q, U, V)$ that fully characterize the polarization state of the synchrotron emission from M87's relativistic jet. The plots are transformed by a one-third power to increase contrast for the displayed quantities. Due to a mean EVPA angle change of approximately $18^{\degree}$, one can see clear distinction between the EVPA angle with no axions (red ticks), and the EVPA angle with axions (cyan ticks). 
\subsubsection{Spatial Dependence}
The Stokes $I$ map (upper left) shows the characteristic bilateral asymmetry from Doppler beaming in the parabolic jet structure, with peak emission concentrated between $50M$ and $150M$ along the projected jet axis. Critically, the total intensity remains unchanged by the presence of the axion field. This confirms the theoretical expectation from the geometric optics formalism—the amplitude equation for $\bar{A}$ contains no axion dependence, meaning axions affect only polarization through the evolution of the polarization vector $\xi^\mu$ rather than the radiated total.
\par
The Stokes $Q$ and $U$ maps (upper right and lower left) encode the linear polarization structure reflecting the jet's helical magnetic field geometry. The $Q$ map displays bilateral asymmetry across the jet axis, while $U$ exhibits an antisymmetric pattern. These complementary structures arise naturally from the magnetic field configuration in our self-similar model. The cyan EVPA ticks show systematic rotation relative to the red ticks, with the magnitude of rotation varying spatially across the jet-photons traversing longer paths through the axion-permeated plasma accumulate larger net phase shifts than those sampling shorter emission path lengths.

\begin{figure*}[t]
    \centering
    \includegraphics[width=\textwidth]{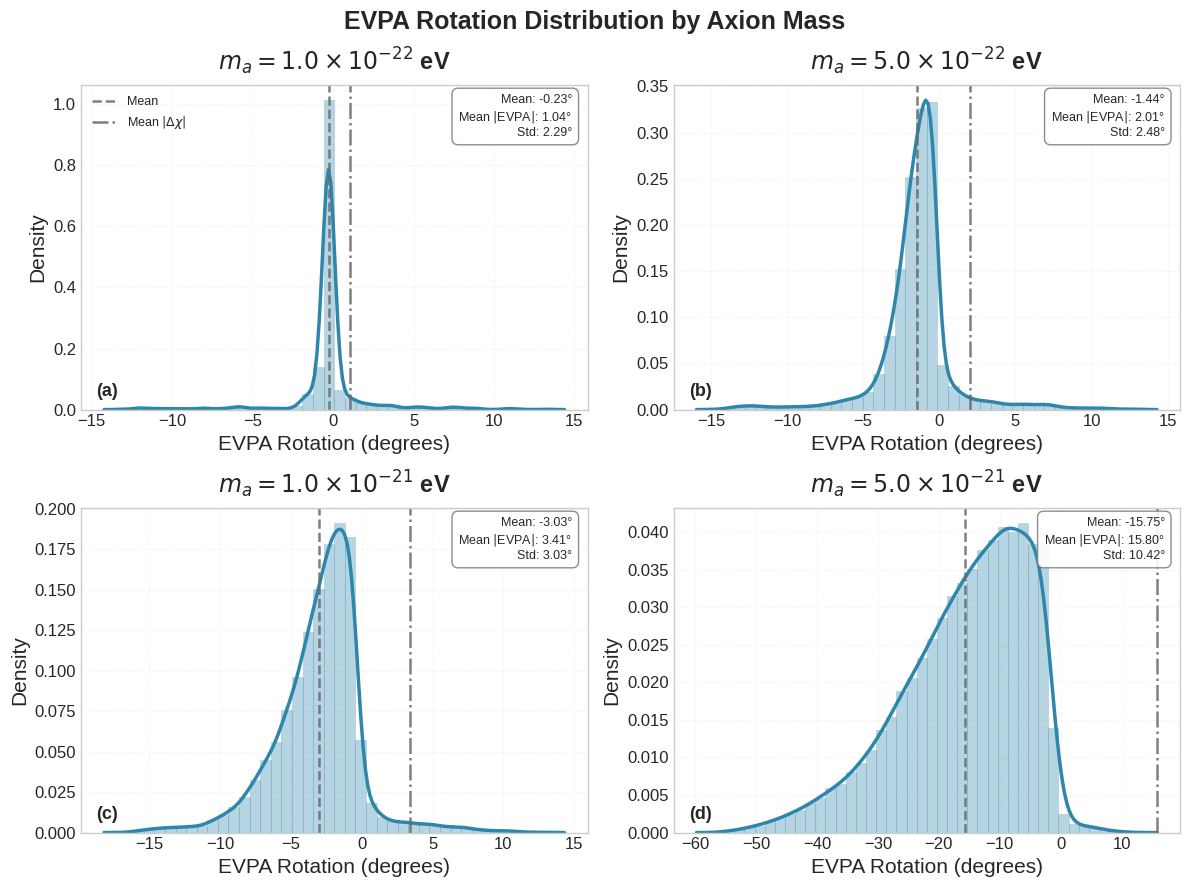}
    \caption{Histogram and PDF distributions for measured $\Delta \chi$ for each mass. Horizontal lines highlight the mean EVPA change with directionality taken into account, and the mean magnitude of EVPA change.}
    \label{fig:distributions}
\end{figure*}

\par
We show an additional example in Fig. \ref{Stokes2} with mass $10^{-22} eV$ and our strongest coupling, $g_{a \gamma}=5 \times 10^{-14} GeV^{-1}$. In this case, we show an example where the EVPA changes is not visible on our Stokes maps, since the mean $\Delta \chi$ across the jet is $< 1^{\circ}$ with very little deviation. 
\subsubsection{Time Dependence}
The map examples shown are a result of the coherent oscillating field reaching its peak amplitude. In reality, time dependent maps would oscillate the EVPA on a characteristic timescale set by the Compton period ($T = \frac{2 \pi \hbar}{m_a c^2}$). For the EVPA maps shown, one would just see this EVPA angle oscillate back and forth on its axis. For the masses considered in this work, the timescales vary from a few days to over a year. These timescales carry direct observational implications. Depending on campaign length and cadence, one may need to carry out a multi-epoch strategy to account for the full oscillation period of the field. These temporal properties of polarized jets occur on timescales that can be probed by EHT black hole dynamical observations %\citep{Satapathy2022}. 

\section{\textbf{Morphology Analysis}}

The imprint of ALPs on the relativistic jet are shown as path-integrated modifications that accumulate along photon trajectories through the jet value, as well as consistency with the azimuthal symmetry of the jet and its magnetic field structure. Different rays traversing the jet produce different cumulative axion-induced rotations. Here, we show this morphology structure showing the effects of various mass and coupling constant values. The analysis in this section reinforces that for our analysis, the masses in the $10^{-21}\rm eV$ regime show these structural properties more comprehensively than $10^{-22}\rm eV$ masses. Because of this, the structural patterns we see are more prevalent for these masses, which we also showcase in this section. Since we calculated the Stokes and EVPA maps for both the axions and no axions case, much of our analysis is focused on the field $\Delta \chi(X,Y)$.  This is not meant to put constraints on the mass or coupling constant values in which the morphology is present, rather, provide intuition on what the effects may look like given certain values. 

\begin{figure*}[t]
    \centering
    \includegraphics[width=\textwidth]{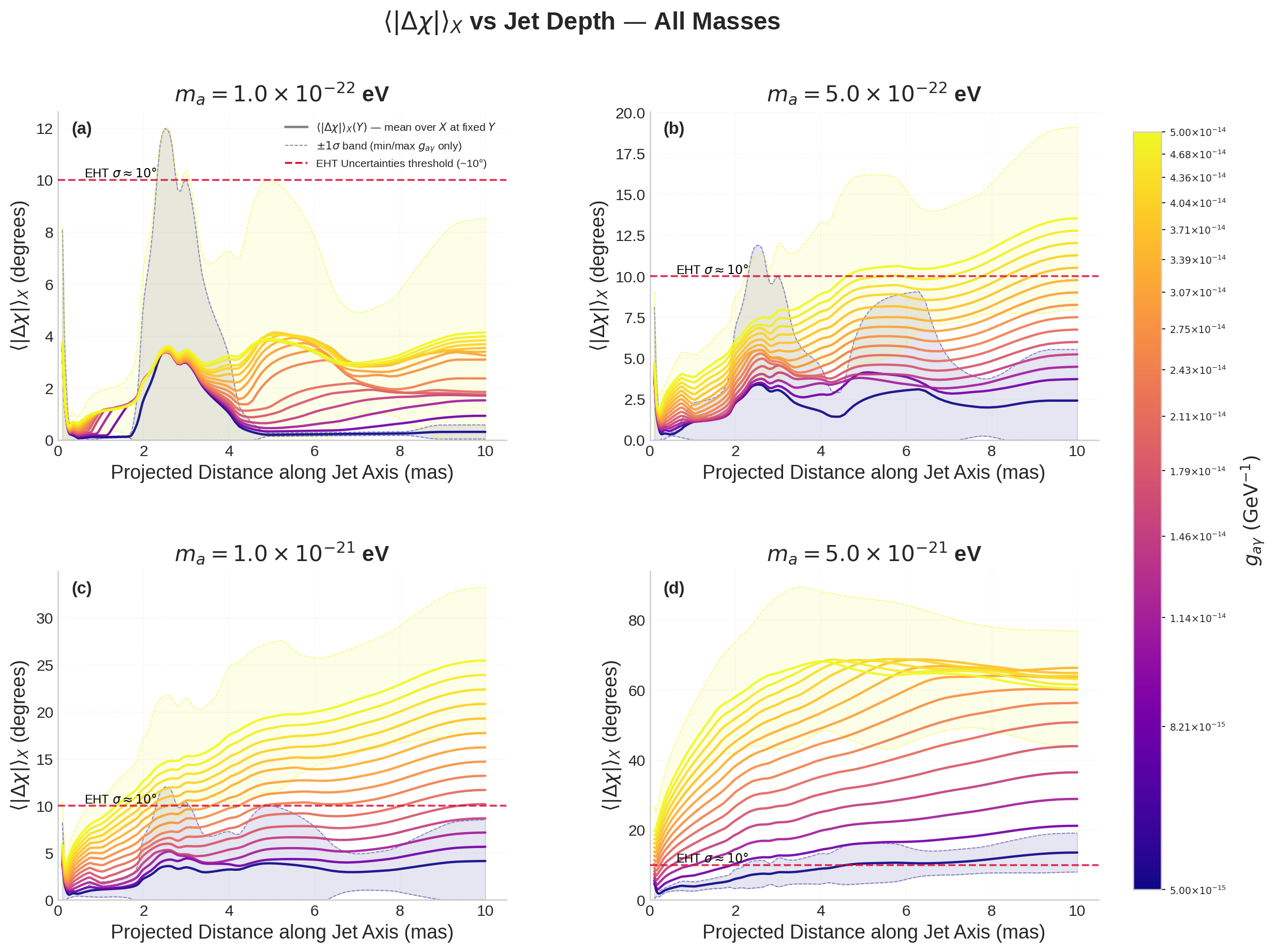}
    \caption{Average change in EVPA values along the jet axis for each mass coupling combination. The red dashed line represents the typical EHT uncertainties for measuring the EVPA angle. }
    \label{fig:GrowthBroad}
\end{figure*}

\subsection{EVPA Signal Patterns}

\begin{figure}[h!]
    \centering
    \includegraphics[width=1.10\linewidth]{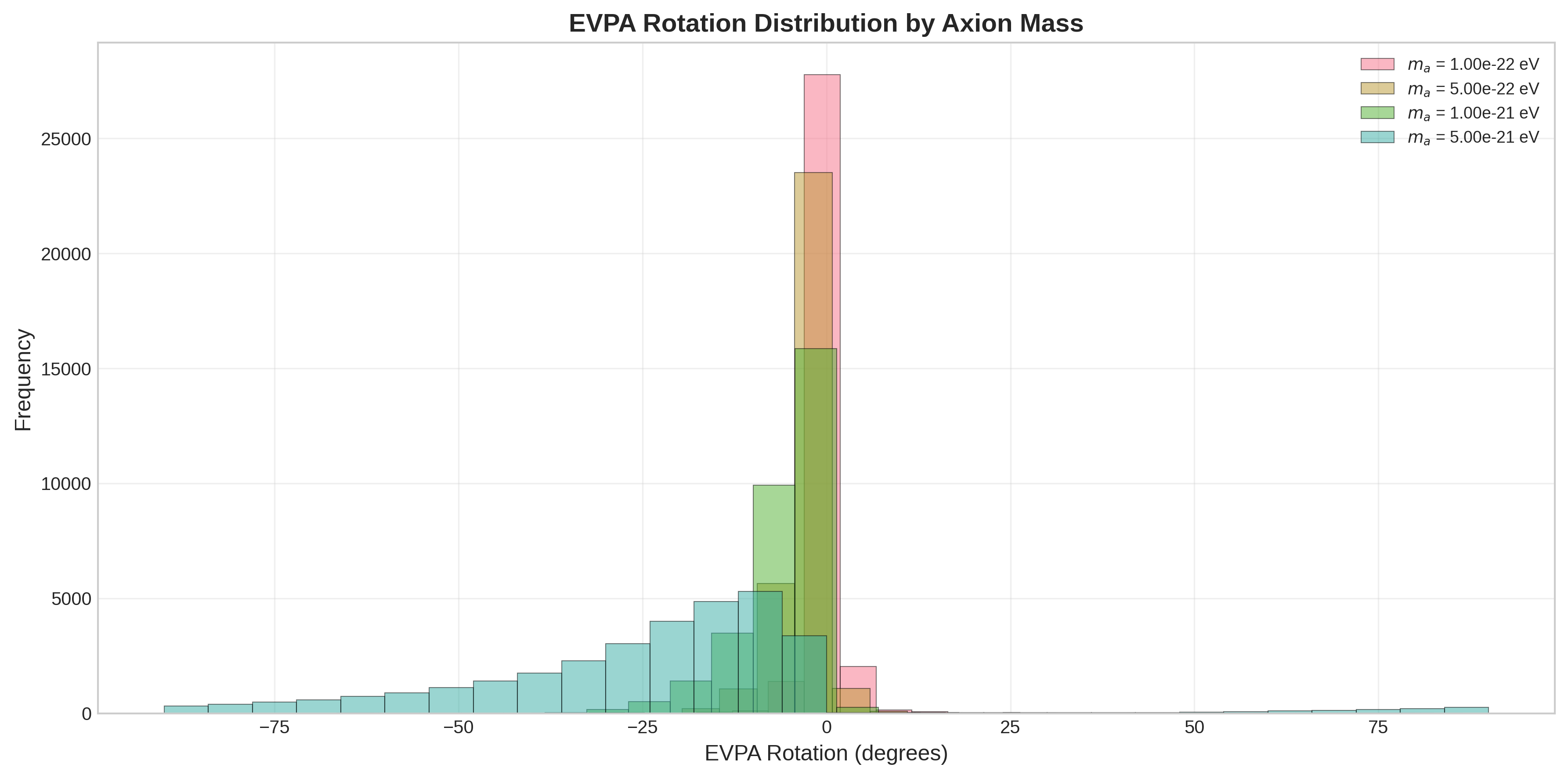}
    \caption{Distribution of EVPA rotation values for each mass.}
    \label{fig:EVPAbyMass}
\end{figure}

For our model, EVPA changes detectable by EHT are more likely to occur in the $\mathcal{O}(10^{-21} \rm eV)$ regime. Fig. \ref{fig:distributions} shows the relative occurrence of each EVPA angles for each mass. Clearly, on average, the two $10^{-21}$ eV masses have a higher likelihood of EVPA changes falling into the detectable regime.  In contrast, the $10^{-22}$ eV masses produce predominantly sub-degree rotations that fall below typical EHT measurement uncertainties. This can be further seen visually in Fig. \ref{fig:EVPAbyMass} which shows the distribution of all EVPA rotations for each mass measured in our simulations. Masses in the $10^{-21} eV$ show larger variance and a higher probability of $\Delta \chi$ lying within the detectable regime, while $10^{-22}\rm eV$ masses are heavily concentrated in the sub-detection regime. 

More importantly, we see a significant relationship between the magnitude of the change in EVPA rotation and the distance along the jet flow. The various growth of rotation curves show the mean and dispersion of EVPA rotation as a function of projected distance along the jet axis. We take the magnitude of the EVPA change since it better shows the relationship. Due to using differential EVPA, rays that traverse a longer path length through the axion-permeated jet accumulate a larger net phase shift. Assuredly, Fig.\ref{fig:GrowthBroad} shows the growth of the EVPA rotation as a function of projected distance along the jet axis for different axion masses and couplings. These figures show that on average, the EVPA rotation grows proportionally with the jet axis up until a certain saturation point.  While the Stokes maps presented in Section 5 are restricted to 1 mas to match the well-constrained emission region of the semi-analytic jet model, the EVPA rotation distributions and growth curves extend to 10 mas to capture the full path-integrated birefringence signal within the soliton core. The two spatial extents probe complementary aspects of the axion signal.

\par
The spread in rotation values at each distance reflects the geometric variation in path lengths across different lines of sight through the jet. Within the uniform-density soliton core, the accumulated EVPA rotation is integrated along the path length through the emitting plasma. Rays along the jet spine traverse longer paths through the emission region and thus accumulate larger rotations, while rays at larger transverse distances exit the optically thick jet earlier and accumulate smaller net rotations. This differential path length effect, combined with the jet's opening geometry, creates the characteristic azimuthal structure in the rotation pattern that we analyze in more detail in Section 6.2. 

\begin{figure}[h]
    \centering
    \includegraphics[width=1.00\linewidth]{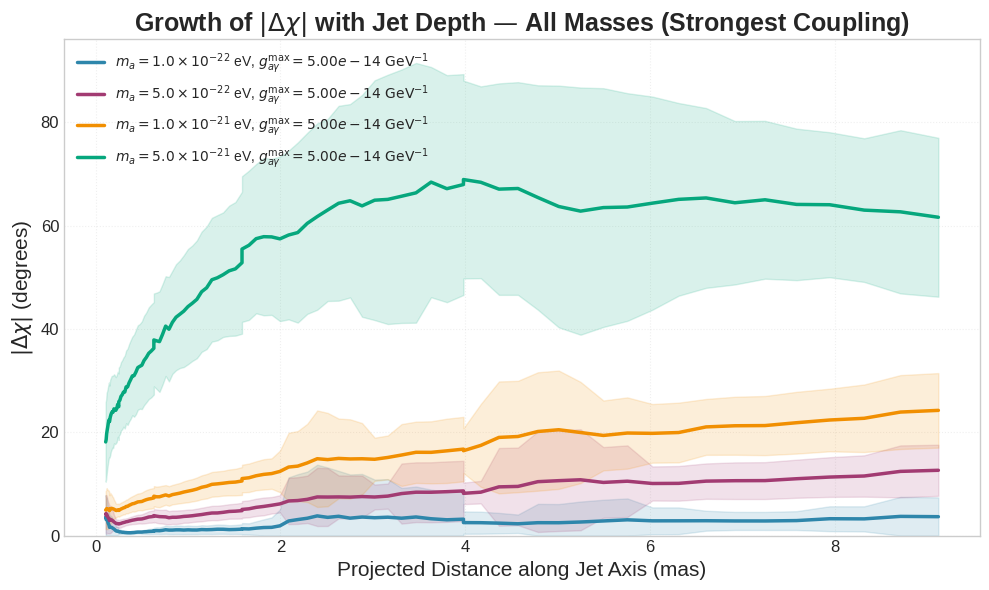}
    \caption{Mean $\Delta \chi$ along the jet axis for each mass and strongest coupling used in our model. The $5 \times 10^{-21} eV$ mass}
    \label{fig:strongestcouplinggrowth}
\end{figure}

\begin{figure*}[t]
    \centering
    \includegraphics[width=\textwidth]{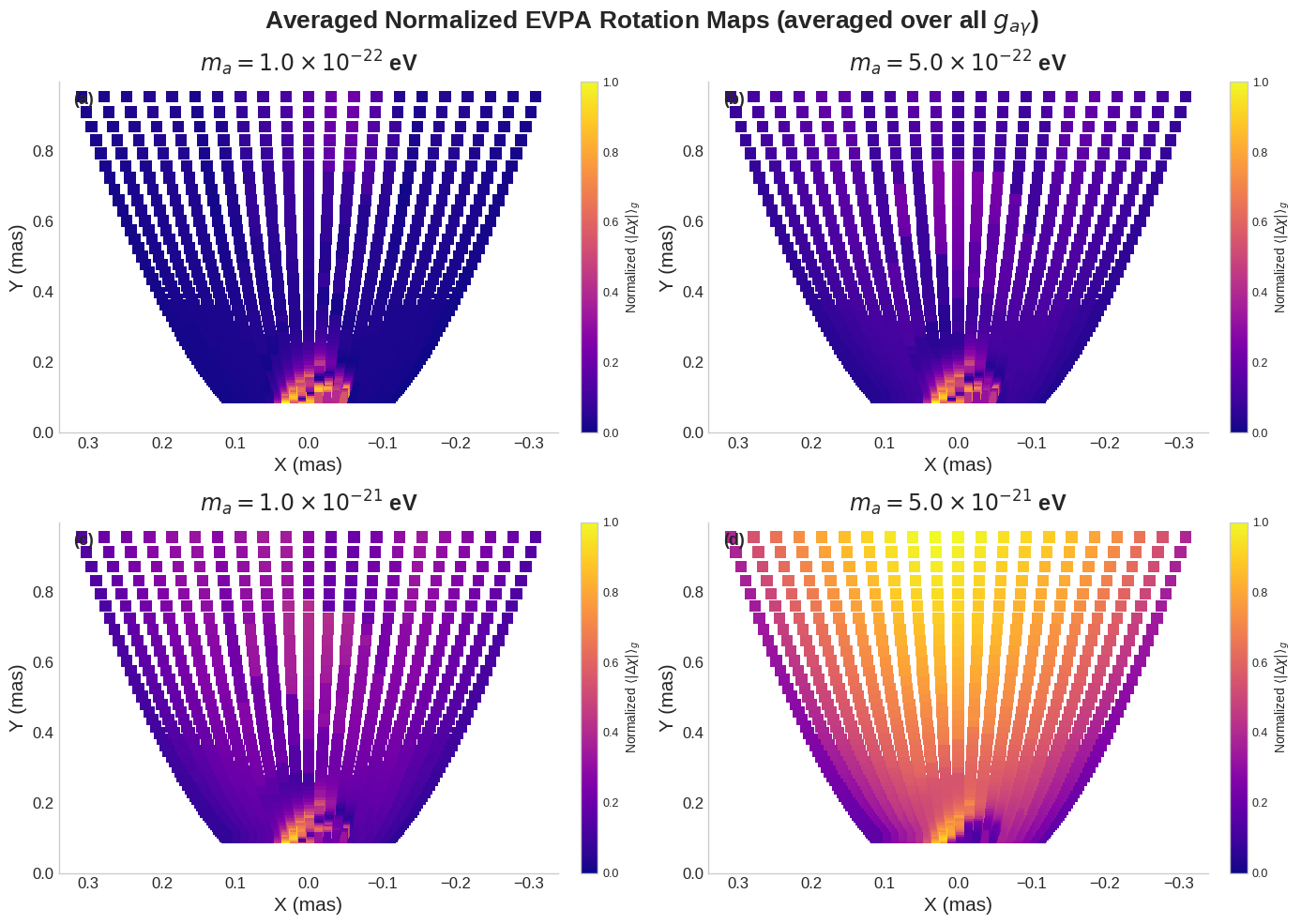}
    \caption{EVPA Rotation Maps 230 GHz showcasing $\Delta \chi (X,Y).$. This is the difference map between EVPA maps with axions and EVPA maps without axions. These should not be used to interpret which regimes are detectable, since this is not made with that scale in mind. }
    \label{fig:RotationMaps}
\end{figure*}

\par
Figure \ref{fig:RotationMaps} is a useful visualization for showing patterns for both and distance along the jet, and the transverse structure (X-direction variation at fixed Y). These plots show the averaged normalized EVPA rotation at each point. For each mass, we averaged over all couplings and then normalized, this way we see the structure that is present for all mass, coupling pairs without the bias of certain pairs being more heavily skewed. Aside from the dense region at the jet base, one can clearly see the tendency for the EVPA rotation magnitude to increase along the jet extent. Additionally, one can clearly see that the magnitude of the $\Delta \chi$ tends to be higher along the central spine ($X \approx 0$) and decreases toward the jet edges (large $|X|$). It should be noted that the growth along Y is not necessarily monotonic. This can be more clearly seen for the stronger couplings of the $5 \times 10^{-21}eV$ masses in Fig. \ref{fig:strongestcouplinggrowth} where the EVPA rotation becomes oscillatory in nature as it saturates. This can also be seen to a slightly lesser degree for the same mass in Fig. \ref{fig:GrowthBroad}.
\par
The detectability of this effect in resolved polarimetric imaging depends not only on the overall amplitude, but also on the spatial coherence of the induced rotation across the image plane. In particular, a smooth, geometry-driven imprint is observationally distinct from a patchy, rapidly de-correlating pattern that could be absorbed into small-scale plasma structure, numerical noise, or beam convolution. Such characteristic spatial scales may also carry information related to the de Broglie wavelength of the axion field. To quantify the characteristic spatial scales of $\Delta\chi(X,Y)$, we employ the second-order structure function.

\begin{figure*}[t]
    \centering
    \includegraphics[width=\textwidth]{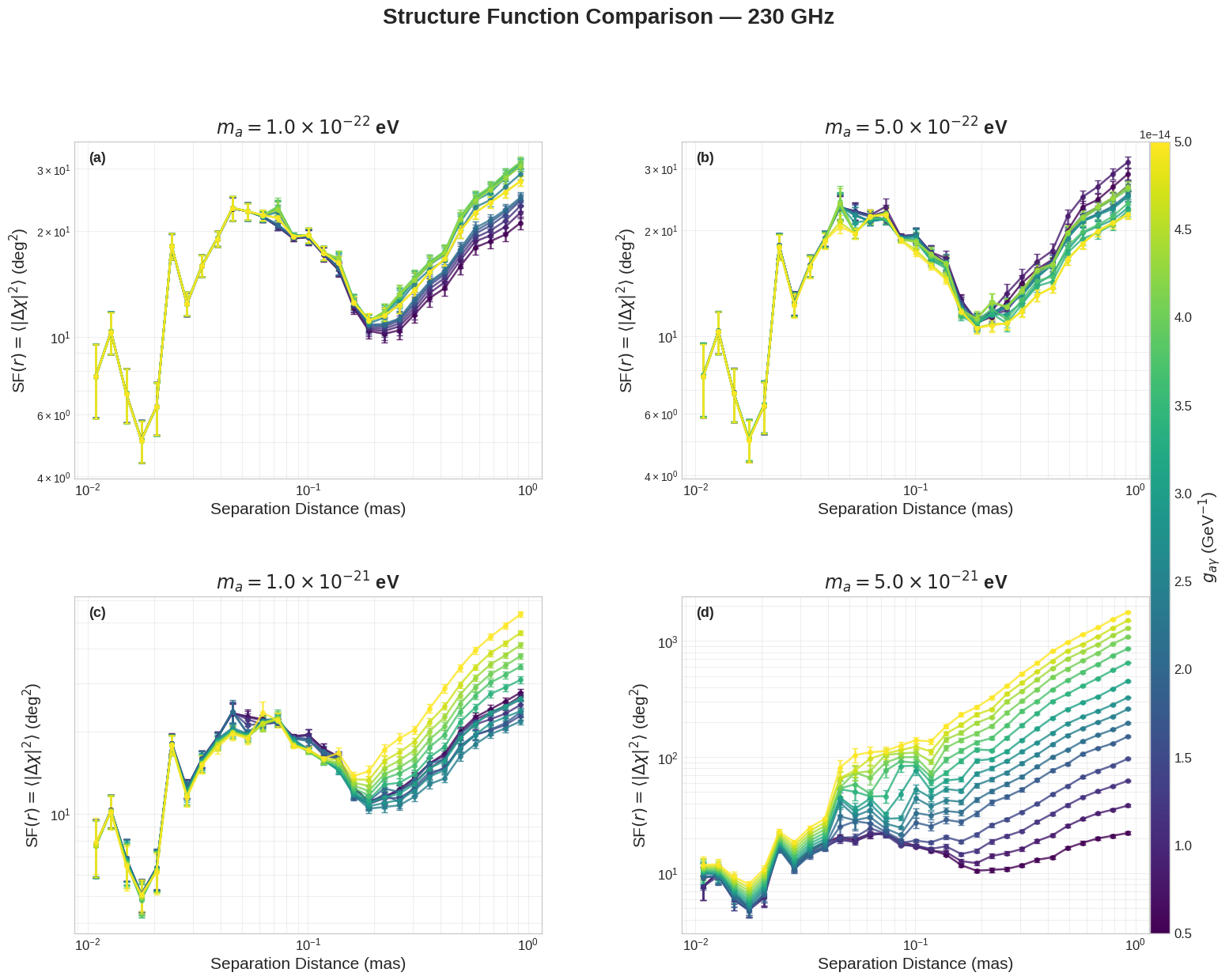}
    \caption{EVPA Rotation Maps 230 GHz. These particular maps are meant to visualize the EVPA patterns of each mass. These should not be used to interpret which regimes are detectable, since this is not made with that scale in mind.}
    \label{fig:StructurFunction}
\end{figure*}

\par
For some measured position $\textbf{x}$ on the image plane, the second-order structure function can be simply defined as,

\begin{equation}
    {S_2(\ell) = \left\langle \left| \Delta \chi(\mathbf{x} + \ell) - \Delta \chi(\mathbf{x}) \right|^2 \right\rangle}
\end{equation}
where $\ell$ is the separation distance. The ALP field has a coherence length $\lambda_{DB} \propto \frac{1}{m_a v}$, meaning a spatial coherence length of order $\ell_c \sim \frac{1}{m_a v}$. For two arrays whose emission points are separated by $\Delta \textbf{x}$, the field values are expected to be strongly correlated as long as $\Delta \textbf{x} \ll \ell_c$. In principle, extending the analysis to much larger separations (comparable to $\ell_c \sim \frac{1}{m_a v}$ would allow the structure function to probe the axion field’s correlation scale and potentially constrain $\lambda_{DB} \sim \frac{2 \pi}{m_a v}$.
\par
Structure function comparisons for each mass and coupling are shown in Fig. \ref{fig:StructurFunction}. As expected, each coupling exhibits the same relative behavior just at different amplitudes. Small spatial separations are more sensitive to pixel-scale discretization, and small scale gradients in the radiative transfer geometry. The large scale variance is more likely geometry dominated rather than axion-dominated.  This is also likely linked to the axion coherence length and its relationship to the beam. 

\subsection{Azimuthal Structure and Symmetry}

Beyond the radial growth of rotation along the jet axis, the axion field induces characteristic azimuthal patterns in the EVPA rotation that provide additional diagnostic signatures. Similarly to the properties discussed in the previous subsection, some aspects of the azimuthal structure are a result of the jet geometry rather than the ALP field, while others also encode the field's characteristic. Even in the absence of spatial gradients in the axion field itself, variations in jet geometry, emissivity weighting, optical depth, and relativistic aberration cause different rays to accumulate different net phase shifts. Thus, even with the axion field being spatially homogeneous in our soliton model, the observed EVPA rotation still acquires azimuthal structure through radiative transfer effects and geometry of the jet. The amplitude and mass-dependent scaling of the azimuthal pattern encode the axion parameters while the shape is geometric. There are also azimuthal effects from sky-plane to emission plane mapping, including the effects of the inclination angles, which introduces mode mixing.

\par
\begin{figure}[h]
    \centering
    \includegraphics[width=0.95\linewidth]{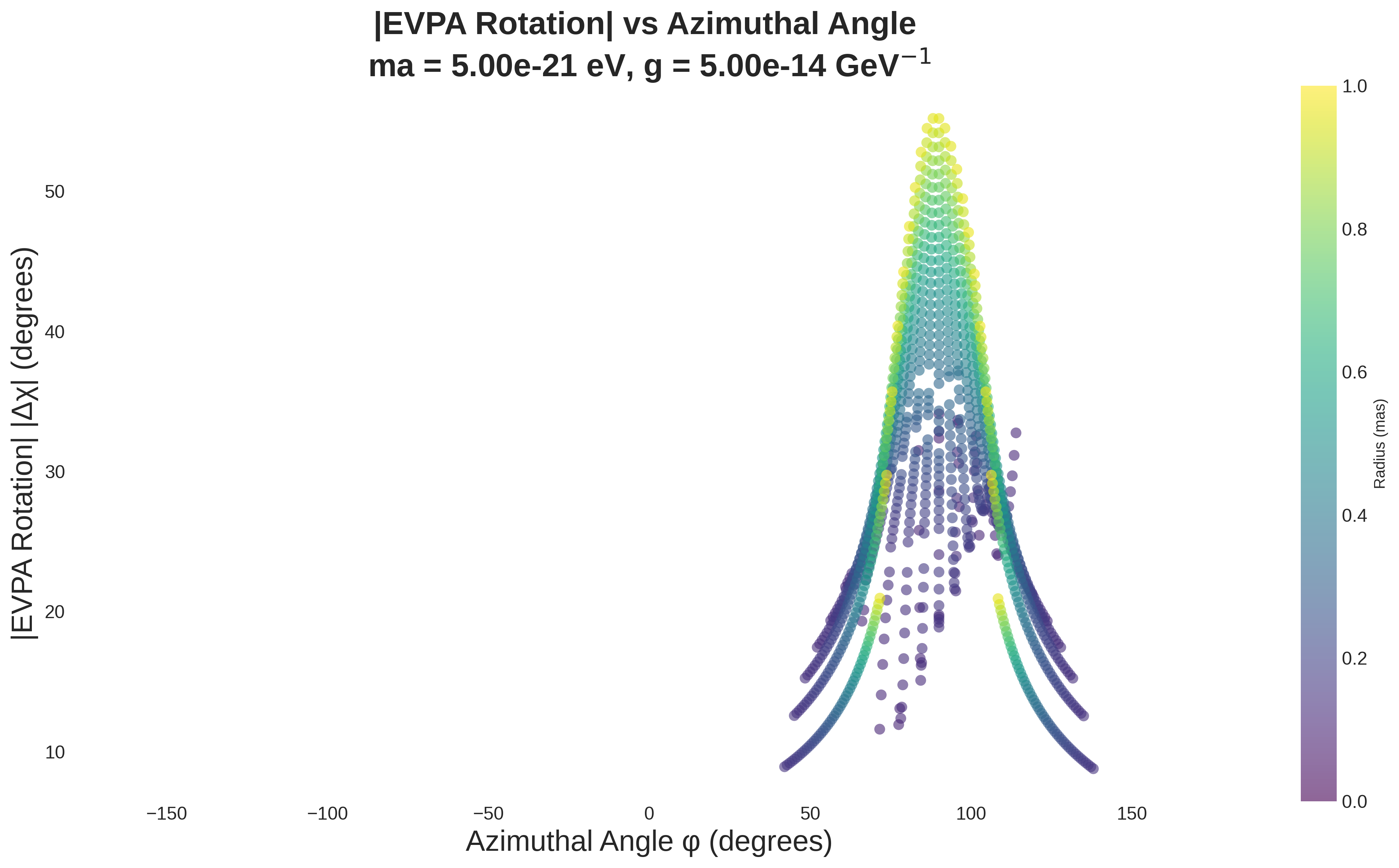}
    \caption{Example graph of the azimuthal structure for $\Delta \chi$ with axion coupling $g=5\cdot10^{-14}GeV^{-1}$ and mass $m_a=5\cdot10^{-21}$ eV.}
    \label{fig:AzimuthalRadius1}
\end{figure}
\begin{figure}
    \centering
    \includegraphics[width=0.95\linewidth]{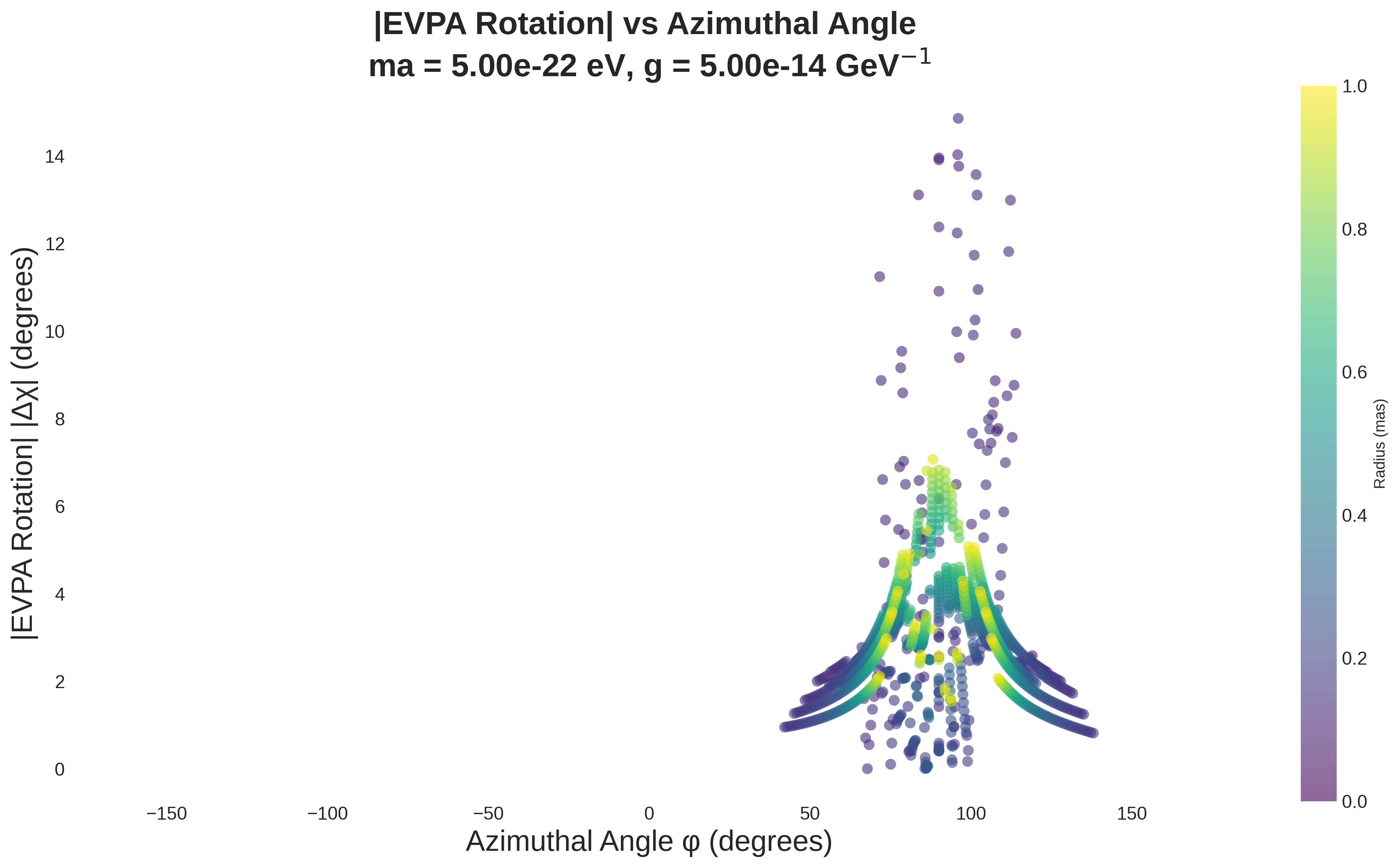}
    \caption{Example graph of the azimuthal structure for $\Delta \chi$ with axion coupling $g=5\cdot10^{-14}GeV^{-1}$ and mass $m_a=5\cdot10^{-22}$ eV. Second example coupling is chosen to explicitly match the first.}
    \label{fig:AzimuthalRadius2}
\end{figure}

Figures \ref{fig:AzimuthalRadius1} and \ref{fig:AzimuthalRadius2} show the per-pixel EVPA rotation magnitude as a function of polar angle around the image center, with color indicating how far from the center the pixel lies. This showcases a relatively symmetric envelope-like shape. Each point represents a pixel in the image plane plotted at its azimuthal angle from the jet axis, so the envelope shows how the maximum rotation amplitude varies with azimuth around the jet.  As the observer sweeps around the image in polar angle, each azimuth corresponds to a different chord through the inclined jet and surrounding soliton core. The narrow peak in EVPA rotation marks the direction in which photons traverse the greatest thickness of emitting plasma within the birefringent medium. Hence, it's a direct imprint of projection of photon paths traveled.  We can see more ``orderly" structure for the larger mass case, and higher scatter at larger radius for the lower mass case.  The offset may be attributed to the inclination angle of M87. For an inclined jet, photons from the receding (far) side travel longer paths through the soliton than those from the approaching (near) side at the same projected distance, breaking the front-back symmetry. The cumulative effects of axion-photon coupling compete with changing jet fluid parameters along the axis such as increasing $v_z$ in the accelerating flow \citep{Anantua2016PhD}, with the degeneracy broken by comparison of the spatiotemporal variation of axion versus plasma induced polarization. The azimuthal signature also provides a potential discriminant between axion-induced birefringence and standard Faraday rotation. While both effects accumulate along the line of sight, Faraday rotation depends on the magnetic field component parallel to the propagation direction, coupling to the helical field geometry of the jet. In contrast, axion birefringence depends on the total traversal through the coherent soliton core and emission weighting, producing azimuthal structure governed by the spatial relationship between the emission region and the spherically symmetric dark matter distribution rather than the magnetic field configuration. Future resolved polarimetric observations of jets at different inclination angles could exploit this distinction to separate axion effects from plasma propagation effects as well as exploit directional dependence seen in the envelope. This conveys a proxy for further azimuthal analysis. 

To analyze symmetry along the jet spine, we perform an independent symmetry 
analysis directly in the observer-plane coordinates $(X, Y)$. The parity decomposition probes left-right symmetry across the jet axis at fixed depth.
At each fixed projected jet distance $Y$, the EVPA rotation profile $\Delta\chi(X)$ can be decomposed into its even (symmetric) and odd (antisymmetric) components by pairing emission points at $+X$ and $-X$ across the jet spine:

\begin{align}
A_0 &= \left\langle \left| \Delta\chi_{\rm even}(X) \right| \right\rangle_X, \\
A_1 &= \left\langle \left| \Delta\chi_{\rm odd}(X) \right| \right\rangle_X, \\
A_2 &= \left\langle 
\left| \Delta\chi_{\rm even}(X) 
      - \left\langle \Delta\chi_{\rm even}(X) \right\rangle_X 
\right| 
\right\rangle_X.
\end{align}

\noindent where $A_0$ is the mirror-symmetric amplitude, $A_1$ is the antisymmetric amplitude, and $A_2$ is a residual quadrupole defined as the even 
component after subtracting its own mean. Each amplitude is averaged over all paired $X$ values at fixed $Y$, and the ratios $A_1/A_0$ and $A_2/A_0$ are then averaged across all coupling constants $g_{a\gamma}$.

\begin{figure*}[p]
    \centering
    \includegraphics[width=\textwidth, height=0.47\textheight, keepaspectratio=false]{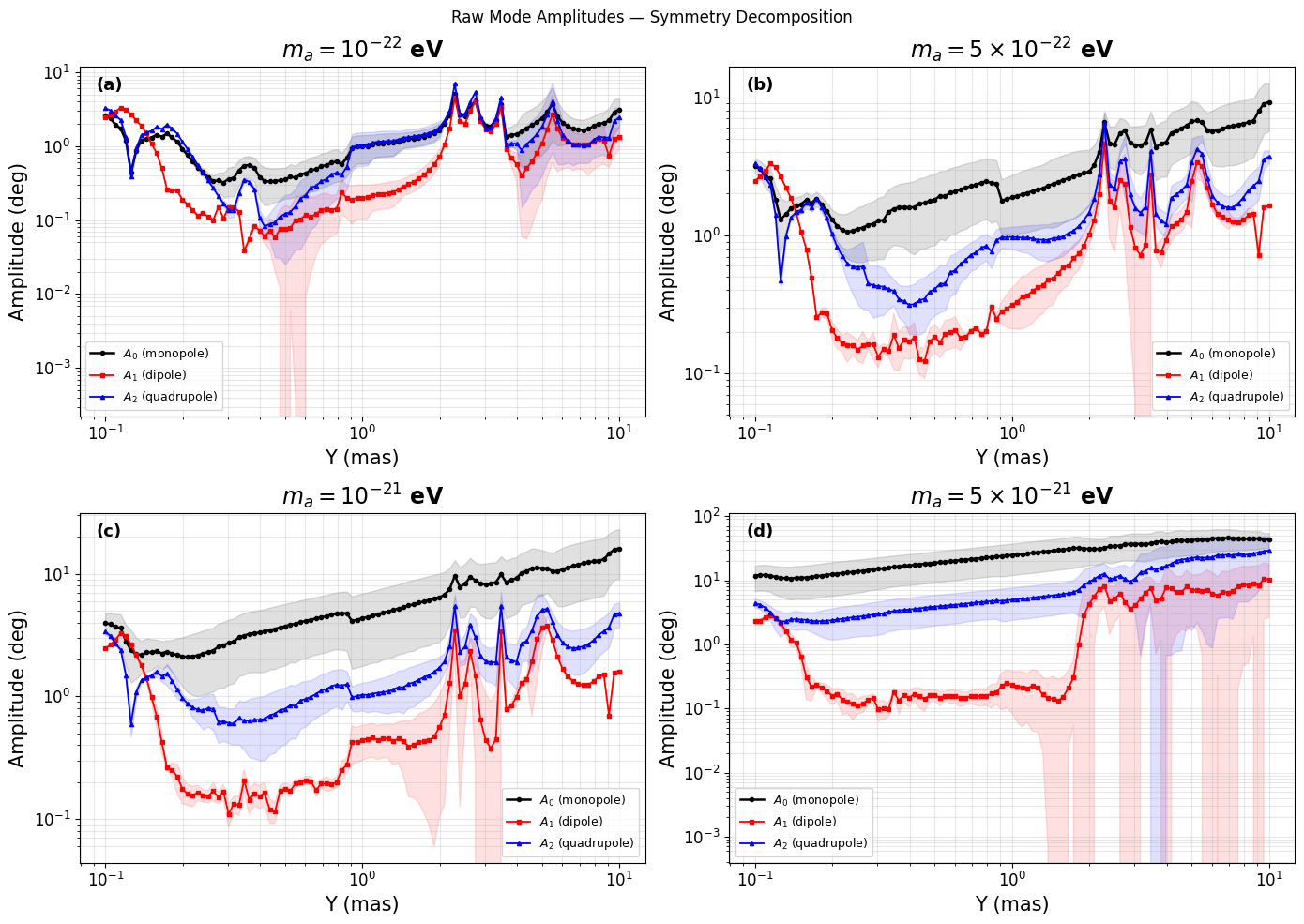}
    \vspace{0.2cm}
    \includegraphics[width=\textwidth, height=0.47\textheight, keepaspectratio=false]{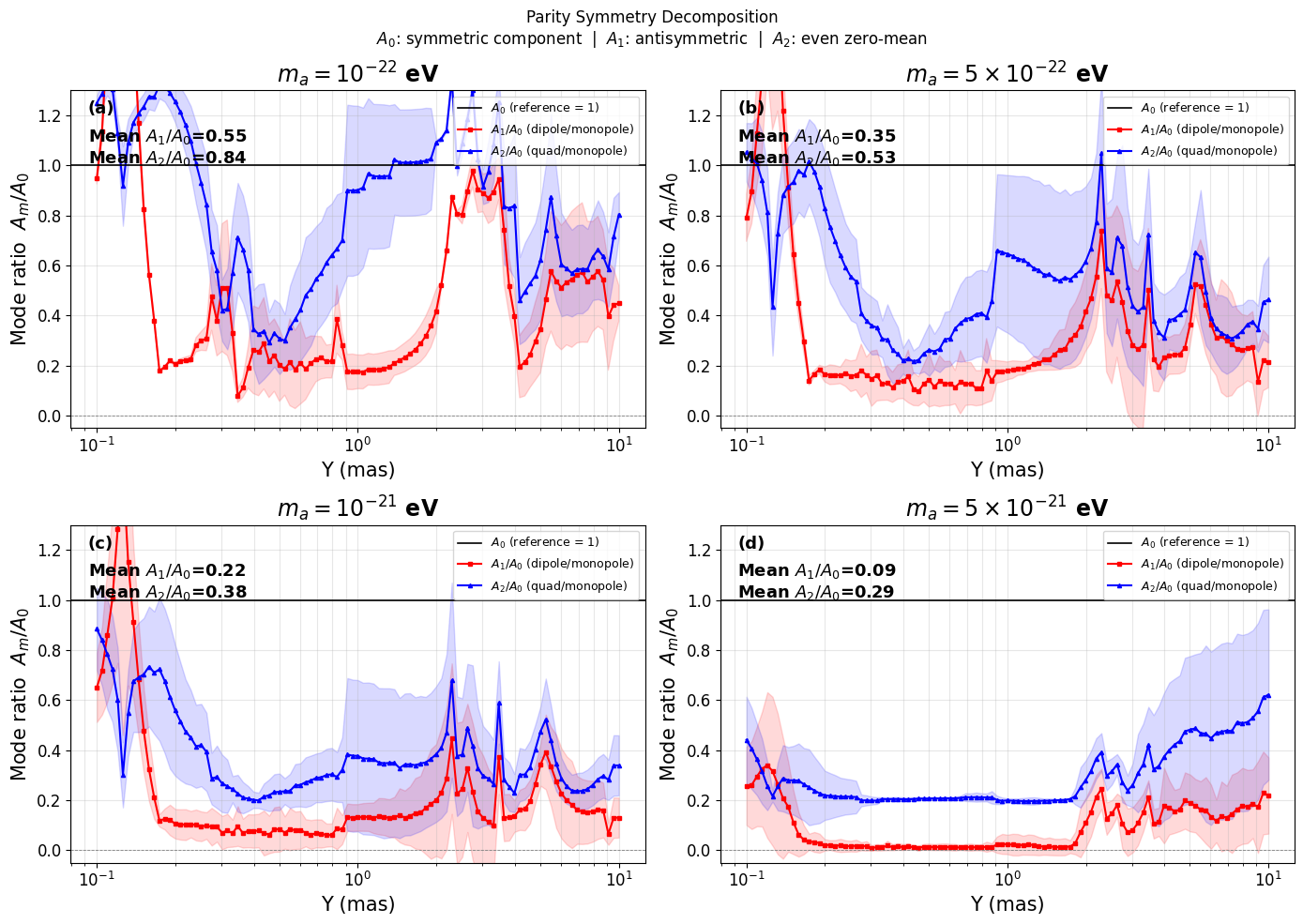}
    \caption{Parity symmetry decomposition of the axion-induced EVPA signal. 
    \textit{Top:} Raw mode amplitudes $A_0$, $A_1$, $A_2$ in degrees 
    (log scale), confirming absolute monopole dominance across all $Y$ 
    and all masses.
    \textit{Bottom:} Mode ratios $A_1/A_0$ (dipole/monopole) and $A_2/A_0$ 
    (quadrupole/monopole) as a function of projected offset $Y$, averaged 
    over all coupling values, for each axion mass. Shaded bands denote 
    $\pm1\sigma$ variation across couplings. The monopole component 
    dominates at all masses, with subdominance increasing with $m_a$.}
    \label{fig:parity_decomposition}
\end{figure*}

Figure~\ref{fig:parity_decomposition} presents both the 
raw amplitudes (top) and the normalized mode ratios (bottom) 
as a function of projected offset $Y$ for each axion mass. 
Several features are immediately apparent. First, the raw 
amplitudes (Fig.~\ref{fig:parity_decomposition}, top) confirm 
that $A_0$ dominates  across all masses and all $Y$, with a few notable exceptions for $\mathcal{O}(10^{-22}eV)$ masses, typically close an order of magnitude over $A_1$. This is consistent with the expectation that a coherent, 
spherically symmetric soliton core produces a predominantly 
even imprint on the transverse EVPA profile. Second, the 
normalized ratios reveal a clear trend with 
axion mass: the mean $A_1/A_0$ ratio decreases from 
$\approx 0.55$ at $m_a = 10^{-22}$~eV to $\approx 0.09$ 
at $m_a = 5\times10^{-21}$~eV, and $A_2/A_0$ decreases 
similarly from $\approx 0.84$ to $\approx 0.29$. Additionally, symmetry analysis could in principle probe coherence length and de Brogile wavelength with jets on large enough scales with deviation of symmetry properties attributable to decorrelation.

\section{Discussion}

\subsection{Morphological Template}

The morphological analyses in this section - the azimuthal mode decomposition, the structure function, and the EVPA rotation growth curves - are performed on the isolated axion-induced EVPA rotation $\Delta \chi(X,Y)$, the pixel-wise difference between EVPA maps computed with and without the axion field. This quantity directly encodes the geometric relationship between the coherent soliton core and the jet's parabolic emission structure, free from the plasma Faraday and synchrotron contributions that are present in the total EVPA. The structure function characterizes the spatial coherence of the axion-induced rotation across the image plane. Azimuthal structure encodes the directionality of photon paths and deeper intrinsic geometric patterns. 
\par
The practical implication is that these morphological features - radial growth of rotation with jet height, azimuthal coherence near the spine, and even-parity dominance - constitute a physical template for what the axion contribution looks like when isolated. In realistic observations, recovering $\Delta \chi$ requires subtracting a well-characterized plasma baseline using multi-frequency observations, where the frequency-independence of axion birefringence distinguishes it from plasma Faraday rotation, and multi-epoch campaigns that track the time-oscillating axion phase. Any spatially coherent EVPA component that persists unchanged across observing bands becomes a candidate axion signal. The morphology identified here provides future analyses with intuition on what templates to construct when probing for searches in polarized images.  

\subsection{Detection Feasibility}
The detection of axion-induced EVPA rotations in relativistic jet polarimetry requires that the predicted signal amplitude exceed the systematic and statistical uncertainties inherent to EHT polarimetric measurements. Current EHT observations at 230 GHz achieve angular resolution of approximately 20 $\mu as$ (FWHM), with recent pilot observations at 345 GHz demonstrating resolution as fine as 19 $\mu as$ and potential for $\sim 13$–$14\,\mu\mathrm{as}$ with the full array. Time-variable Faraday 
rotation from external screens introduces additional EVPA corruption of $\sim \pm 15\degree$ (Event Horizon Telescope Collaboration et al. 2024). Taken together, these systematics establish an effective detection threshold of approximately $3–5\degree$, which is generally considered a conservative estimate. The next-generation Event Horizon Telescope (ngEHT) is expected to dramatically expand sensitivity to subtle EVPA signatures, potentially being able to identify axion-induced EVPA changes from the previously discussed sub-threshold regimes. With its enhanced bandwidth, the ngEHT may be able to detect EVPA fluctuations as small as $0.05\degree$ \citep{ayzenberg2023fundamentalphysicsopportunitiesnextgeneration}. The EHT has revealed the highest resolution polarization structure for jetted AGN in M87 \cite{EHTM87PaperVII2021} and OJ 287 \cite{Gomez2026}. Horizon-scale M87 observations of vertical and poloidal near horizon field geometry support the interpretation of M87 as a Blandford-Znajek jet. $\rm OJ\ 287$ helical jet structure support the same interpretation, making our Blandford-Znajek jet model in this work a good starting point for comparison with observation. The numerical EVPA measurements of $199^\circ \pm 11^\circ$ and  $244^\circ \pm 10^\circ$  on April 5, 2017 and April 11, 2017 have uncertainties that are exceeded by axion induced polarization swings in several of our models. 
\par
For a mass of $5 \times 10^{-21} eV$, the majority of couplings showed measured EVPA rotation distributions where the majority of values exceed $10^{\circ}$ which is well within the EHT detectability range. Many EVPA angle measurements for $m_a = 10^{-21} eV$ fall into this range for the stronger couplings. Our masses of $\mathcal{O}(10^{-22} eV)$ fall well below these these thresholds with the exception of the most extreme cases for $5 \times 10^{-22} eV$. However, these regimes may be detectable with the ngEHT. We stress that this does not put any parameter constraints for axion detection with this model since this an idealized case. 
\par
The EHT collaboration has established a powerful precedent for extracting physical constraints from polarization morphology rather than absolute EVPA values. Introduced in \citep{Palumbo_2020}, the complex coefficient $\beta_2$ has proven highly effective at distinguishing accretion states, encoding information about the coherence of the polarization spiral and its handedness. 
\cite{2023ApJ...950...38E} investigated the origin of the twisty patterns in the linear polarization and showed that this twisty morphology is dictated by the magnetic field structure in the emitting region. Furthermore, \cite{2023ApJ...950...38E} showed that, the dependence of the twisty pattern on the BH spin is attributed to variations in the magnetic field geometry occurring owing to the frame dragging. 
\cite{Chael_2023} subsequently demonstrated that $\beta_2$ is strongly coupled to black hole spin and encodes the direction of electromagnetic energy flow near the horizon, providing constraints on spin and energy extraction. The recent detection of EVPA helicity reversal between the 2017 and 2021 M87 campaigns demonstrates that the EHT can robustly identify morphological changes in polarization structure. When analyzing these changes for axions, one must stress the frequency-independent nature of axion-induced polarization, such that it is clear the structure changes are unique to axions rather than induced from the Faraday plasma.

\subsection{Caveats}
Our analysis makes several simplifying assumptions that warrant critical examination. Real observations will face complications from astrophysical variability, uncertain model parameters, and instrumental effects that could obscure or mimic axion signatures. The semi-analytic jet model employed in this work assumes stationarity, axisymmetry, and self-similarity. While these assumptions enable efficient parameter space and morphology exploration, and are computationally inexpensive, real jets violate all three. 
\par 
Our model uses the Blandford-Znajek force-free field configuration with predominantly poloidal fields near the axis and return currents in the sheath. Real jets have more complex field topologies with turbulent tangles, reconnection sites, and episodic flux eruptions. These alter the plasma Faraday rotation profile which must be accurately modeled to isolate the axion contribution more successfully. The assumption of axisymmetry neglects 3D instabilities such as the current-driven kink mode, which can produce helical jet distortions. These introduce azimuthal EVPA variations that could be degenerate with axion effects. High-resolution GRMHD simulations will be essential for quantifying these systematics.
\par
GRMHD simulations show that jets exhibit strong magnetohydrodynamic turbulence on dynamical timescales on the order of days for M87. This produces stochastic EVPA fluctuations  that could mask or be confused with axion signatures. Critically, turbulent EVPA variations are spatially incoherent on scales below the turbulent coherence length.
\par
Additionally, the signal can be heavily suppressed due to washout. Consider the simplified radiative transfer equation for linear polarization in the presence of axion-induced birefringence:

$$\frac{d(Q + iU)}{ds} = j_Q + i j_U - i 2 g_{a\gamma\gamma} \frac{da}{ds}(Q + iU)$$

where $j_Q$ and $j_U$ are the linearly polarized emissivities and $s$ is the affine parameter along the photon geodesic. The solution to this equation is

$$Q(s_f) + iU(s_f) = \int_{s_i}^{s_f} e^{i 2 g_{a\gamma}[a(s_f) - a(s)]} \left(j_Q(s) + i j_U(s)\right) ds.$$

For optically thin accretion flows, photons reaching the observer originate from different spatial points along the line of sight, each experiencing a different local axion field value due to the coherent oscillation of the axion cloud. When integrated over a finite radiation length $s_r$, the oscillating phase factor $e^{i 2 g_{a\gamma\gamma}[a(s_f) - a(s)]}$ leads to destructive interference, suppressing the EVPA oscillation amplitude. This washout becomes significant when the radiation length is comparable to the axion Compton wavelength, $s_r \sim 2\pi\lambda_c$. For a RIAF with thickness parameter $H \equiv h/R$, the condition $\rho H \ll 2\pi\lambda_c$ must be satisfied to preserve the signal.
\par
When discussing the structure function, we mentioned that the coupling constant values did not change the behavior, but the amplitude of the structure function values. For our values, we saw that for small separation distances, the amplitudes were ordered in accordance to the value of $g_{a \gamma}$ (i.e. higher amplitude for higher $g_{a \gamma}$). However, the observed morphology of the structure functions may change in regimes where EVPA wrapping and depolarization become important.
\subsection{Future Directions}
The results presented here establish relativistic jet polarimetry as a viable morphological probe of axion-induced birefringence under the simplified soliton-core assumption. Several extensions are required to transition from morphological intuition to a fully predictive and data-driven inference framework.
\par
In conjunction with the majority of axion literature related to the EHT, we will be investigating axion effects using the superradiance model of axions as opposed to the soliton core model. This especially becomes essential in mass regimes where the superradiance condition is met. A field of bosonic particles can extract angular rotation from a spinning black holes when the Compton wavelength is comparable to the black hole size, causing field  \citep{Brito_2020}. Since the axion field is azimuthally dependent in the superradiance, this adds additional complexity for the azimuthal modes.  Additionally, jets tend to align more with the spin axis whereas accretion disk emission tends to align with the equatorial plane. Unless the jet has a very large opening angle, there will be a supression of EVPA angle changes since superradiant axions in the dominant modes mainly concentrate near the equatorial region. While studying superradiance will add additional layers of complexity to both the geometry of the signal and its feasibility for detection, this also means we have multiple dark matter models to compare with the EHT for VLBI analysis. Moreover, the soliton core model will continue to be used albeit with more accurate microphysics. With the updated soliton core, we will also be evolving the axion field as a Schrodinger Poisson system. This will likely cause a supression of the induced EVPA angle changes if the soliton survives. 
\par
Patently, various axion models will be studied using GRMHD simulations to incorporate more realistic physics and the astrophysical uncertainties that naturally arise in this environment. A complete treatment would couple the oscillating axion background to time-dependent GRMHD, enabling predictions for both the secular EVPA gradient computed here and the time-variable component that could be extracted from multi-epoch observations. Additionally, superradiant axion clouds localized near the horizon could produce substantially larger signals in the photon ring itself. End-to-end simulations incorporating realistic EHT (u,v)-coverage, thermal noise, and Faraday rotation from the turbulent accretion flow will be essential to quantify detection prospects and optimize observing strategies.
\par
In many future works, we will propagate the axion-rotated Stokes maps through an EHT-like measurement operator (beam/uv sampling and imaging priors) to more accurately quantify signal survival. The present analysis assumes access to fully resolved polarization maps, but real EHT observations sample only discrete points in the $\rm(u,v)$ plane; it remains to be demonstrated that the radial EVPA gradient survives this filtering and can be distinguished from imaging artifacts. A quantitative inference framework must be constructed to compare model predictions directly with existing EHT data. This includes likelihood-based fitting in both image and visibility space, incorporation of multi-frequency constraints to exploit the frequency-independence of axion birefringence, and multi-epoch analysis to separate coherent axion oscillations from stochastic turbulent EVPA fluctuations. Forecasting analyses for the ngEHT, with its expanded bandwidth and improved sensitivity, will determine the parameter regions in which currently sub-threshold signatures may become accessible.
\par
Many observable quantities of interest for the EHT collaboration naturally occur in visibility space. The visibility space 
naturally be written as follows: 

\begin{equation}
    V(u, \theta)=\iint I(\rho, \phi) e^{-2 i \pi \rho u \cos (\theta-\phi)} \rho d \rho d \phi
\end{equation}

where $I$ is intensity in the image space and u is spatial frequency of the Fourier component. One such observable of interest introduced in \citep{emami2023ebcorrelationresolvedpolarizedimages, 2025arXiv250404695E}, characterizes the correlations between the $E$ and $B$ modes of the linear polarization in a visibility space. 

\begin{equation}
    \rho_{\mathcal{E} \mathcal{B}}(u, v) \equiv \frac{E(u, v) B^*(u, v)}{\sqrt{E(u, v) E^*(u, v)} \sqrt{B(u, v) B^*(u, v)}}
\end{equation}

Generally speaking, visibility space for axion detection is rather unexplored for the EHT, especially in the case of jets. 
\par
A deeper exploration of the time-dependent structure of axion-induced birefringence is warranted. An important next step is to systematically predict the polarization morphology associated with each $(m_a, g_{a \gamma})$, combination and to determine the observational timescales over which these signatures would manifest in resolved jet polarimetry.

\section{Conclusion}
Relativistic jets represent an underexplored but promising laboratory for searching for ultralight axion-like particles (ALPs) via polarimetric imaging. In this work, we have demonstrated, that axion birefringence leaves spatially structured, morphologically distinctive imprints on the polarized emission of M87's relativistic jet — imprints that are in principle separable from standard plasma Faraday effects.
\par
Soliton cores can form in galactic centers due to the balance between gravitational pressure and quantum interactions. In the presence of an axion cloud, the EVPA angle of linearly polarized photons oscillates periodically. We included a solitonic axion background for radiative transfer in a semi-analytic model based on a MAD force free Blandford-Znajek jet simulation (based on M87{\color{black}*} parameters). We input a sub-equipartition emission model for synchrotron emission. Our simulations demonstrate that axion birefringence produces spatially coherent EVPA perturbations rather than pixel-level noise. These morphological diagnostics—correlation of rotation amplitude with jet height, azimuthal coherence of the rotation pattern, and frequency-invariance—constitute a multi-dimensional discriminant that can be applied using similar analysis frameworks developed for EHT model comparison, without requiring absolute EVPA calibration at the few-degree level.
\par
The frequency-independent nature of axion birefringence provides an additional independent axis: multi-frequency EHT campaigns at 86, 230, and 345 GHz can in principle isolate the axion contribution by identifying EVPA structure that does not rotate with frequency, as a plasma-Faraday contribution would. This frequency-tagging strategy is complementary to the morphological tests described above and should be incorporated into future observational analysis pipelines.
\par
Several important caveats bound the present analysis.The semi-analytic, stationary jet model neglects magnetohydrodynamic turbulence, kink instabilities, and episodic flux eruptions present in real jets — all of which introduce stochastic EVPA fluctuations that could mask or mimic axion signatures on short spatial scales. The soliton core model itself may not survive intact near M87's supermassive black hole for masses $m_a > 10^{-22} eV$ as the presence of the strong gravitational potential from the SMBH shrinks or destroys the core; throughout this work the soliton density profile is therefore treated as a phenomenological proxy for a compact ALP overdensity rather than a guaranteed cosmological configuration. Finally, the analysis assumes fully resolved polarization maps, while real EHT observations sample a discrete (u,v)-plane; it remains to be demonstrated that the radial EVPA gradient identified here survives beam convolution, interferometric filtering, and imaging reconstruction artifacts.

\par
While the morphological signatures identified here are intentionally broader than the specific observables EHT researchers would prioritize in a dedicated search, this simple approach affords considerable physical intuition about how axion-induced birefringence manifests across the jet structure. This intuition becomes especially valuable as the field moves toward more sophisticated models, where having a clear baseline understanding of the underlying morphology will be essential for interpreting the richer, more complex signals that arise from realistic jet physics and axion configurations. The present work thus serves as a stepping stone, establishing jet polarimetry as a promising laboratory for ultralight axion searches while looking forward to making both the jet model and the underlying axion assumptions more sophisticated in future analyses, with the morphological predictions presented here serving as an initial set of targets for forthcoming observational and simulation studies.

\begin{acknowledgements}
This work was supported by the SCEECS Research Scholars. This work was supported by a grant from the Simons Foundation (00001470, R.A., D.J., N.L.). We would like to thank Yifan Chen, Mustafa Amin, Andrew Long, Dom Pesce, and Roger Blandford for the fruitful conversations regarding this work.
\end{acknowledgements}
\bibliography{bibliography}

\end{document}